\documentclass[times, review, 10pt]{elsarticle}

\usepackage{setspace} 

\usepackage{hyperref}
\usepackage[utf8]{inputenc}
\usepackage[T1]{fontenc}
\usepackage{color,soul}
\usepackage{times}
\usepackage{subfigure}
\usepackage{amssymb} 
\usepackage[nointegrals]{wasysym} 
\usepackage{xcolor}
\usepackage{epsfig}
\usepackage{graphicx}
\usepackage{multicol,multirow}
\usepackage{booktabs}
\usepackage{bigstrut}
\usepackage{floatrow}
\usepackage{rotating}
\usepackage{adjustbox}
\usepackage{mathtools}
\usepackage{subfigure}
\usepackage{longtable}
\usepackage{array}
\usepackage{amsmath,amssymb,amsfonts}
\usepackage{algorithmic}
\usepackage{graphicx}

\usepackage{graphicx,subfigure}
\graphicspath{ {./images/} }
\usepackage{booktabs}
\usepackage{adjustbox}
\usepackage{tablefootnote}
\usepackage{multirow}
\usepackage{textcomp}
\usepackage{xcolor}

\journal{arXiv}

\begin{document}

\begin{frontmatter}



\title{TAFM-Net: A Novel Approach to Skin Lesion Segmentation Using Transformer Attention and Focal Modulation}


\author[inst1]{Tariq M Khan}
\author[inst1]{Dawn Lin}
\author[inst2]{Shahzaib Iqbal}
\author[inst1]{Erik Meijering}
\affiliation[inst1]{
    organization={School of Computer Science and Engineering, UNSW},
    city={Sydney},
    country={Australia}
}
\affiliation[inst2]{
    organization={Department of Computing, Abasyn University Islamabad Campus},
    city={Islamabad},
    country={Pakistan}
}

\begin{abstract}

Incorporating modern computer vision techniques into clinical protocols shows promise in improving skin lesion segmentation. The U-Net architecture has been a key model in this area, iteratively improved to address challenges arising from the heterogeneity of dermatologic images due to varying clinical settings, lighting, patient attributes, and hair density. To further improve skin lesion segmentation, we developed TAFM-Net, an innovative model leveraging self-adaptive transformer attention (TA) coupled with focal modulation (FM). Our model integrates an EfficientNetV2B1 encoder, which employs TA to enhance spatial and channel-related saliency, while a densely connected decoder integrates FM within skip connections, enhancing feature emphasis, segmentation performance, and interpretability crucial for medical image analysis. A novel dynamic loss function amalgamates region and boundary information, guiding effective model training. Our model achieves competitive performance, with Jaccard coefficients of 93.64\%, 86.88\% and 92.88\% in the ISIC2016, ISIC2017 and ISIC2018 datasets, respectively, demonstrating its potential in real-world scenarios.
\end{abstract}



\begin{keyword}
Dermatological image analysis \sep skin lesion segmentation \sep focal modulation \sep transformer attention 
\end{keyword}

\end{frontmatter}


\section{Introduction}

The increasing incidence and mortality rates associated with skin cancer make it a major public health concern. Melanoma is the most lethal form among all skin cancers. Early detection of skin cancer is crucial for the treatment and survival of patients. The earlier skin lesions are detected, the greater the chances that patients will receive appropriate treatment and have a higher probability of recovery. Early detection of melanoma, which manifests itself as pigmented lesions on the surface of the skin, is feasible by using expert visual inspection. However, dermatologists find the diagnosis of skin cancer difficult due to the abundance of skin lesions and the difficulty in differentiating between benign and malignant lesions.

Pigmentation lesions, which appear on the skin's surface, provide the opportunity for early detection of melanoma through expert visual examination. Dermatoscopy, a noninvasive diagnostic imaging technique utilised in dermatology, facilitates the analysis of skin pigmentation. It improves visualisation of deeper skin layers by eliminating surface reflections, allowing dermatologists to identify early-stage melanoma that may be undetectable with the naked eye. Previous research has shown that dermatoscopy has a higher diagnostic accuracy than conventional photography. However, manual examinations can pose specific difficulties in clinical settings as a result of the intricate nature of the lesions and the increased volume of dermoscopic images. These assessments can be subjective, time-consuming, and difficult to reproduce. Furthermore, the interpretation of dermoscopic images of melanoma can be challenging even for experienced dermatologists, particularly in complex cases. For these reasons, the advancement of computer-assisted diagnosis (CAD) has generated significant attention, as it has the potential to help dermatologists in their clinical evaluations \cite{khan2024esdmr,khan2022leveraging,khan2022t,khan2022mkis,naqvi2023glan,iqbal2023ldmres,qayyum2023two,javed2024advancing}.

The ability to autonomously segment lesions from dermoscopic images is a crucial aspect of a CAD system. Accurate segmentation allows subsequent automatic analysis to concentrate specifically on these regions \cite{iqbal2022recent, soomro2016automatic,khan2019boosting,khan2020shallow,arsalan2022prompt,khan2022neural,khan2023retinal}. However, designing an automated segmentation algorithm is a significant challenge \cite{khan2020exploiting,khan2020semantically,khan2021residual,khan2021rc}, due to potentially large variations in the size, shape, and color of the lesions, variations in the pixel values and textures within the lesions and surrounding normal skin, and the presence of fuzzy lesions boundaries resulting from insignificant pixel distinctions between the lesion and the skin\cite{naveed2024pca,naveed2024ra,iqbal2024tesl,naveed2024ad}.

Recently, many models using deep convolutional neural networks (CNNs) have been introduced for skin lesion segmentation. A popular architecture for this task is the well-known U-Net \cite{ronneberger2015}. To achieve more accurate segmentation, the use of both global and local information is key. Ghafoorian \textit{et al.~}\cite{ghafoorian2016} introduced a multistream network that uses multiscale encoders to collect context data at different scales. However, the depth of the network is insufficient to reveal unique features. By including batch normalisation and a residual module, we can effectively address the issues of vanishing or exploding gradients and network degeneration, enabling us to construct deeper networks. Yu \textit{et al.~}\cite{yu2017} showed that a deeper network can capture more comprehensive and distinctive features in the segmentation of skin lesions. However, the study focused mainly on local contexts and overlooked global characteristics, which hindered the achievement of more favourable results with a deeper network.

\begin{figure*}[!t]
    \centering
    \includegraphics[width=\textwidth]{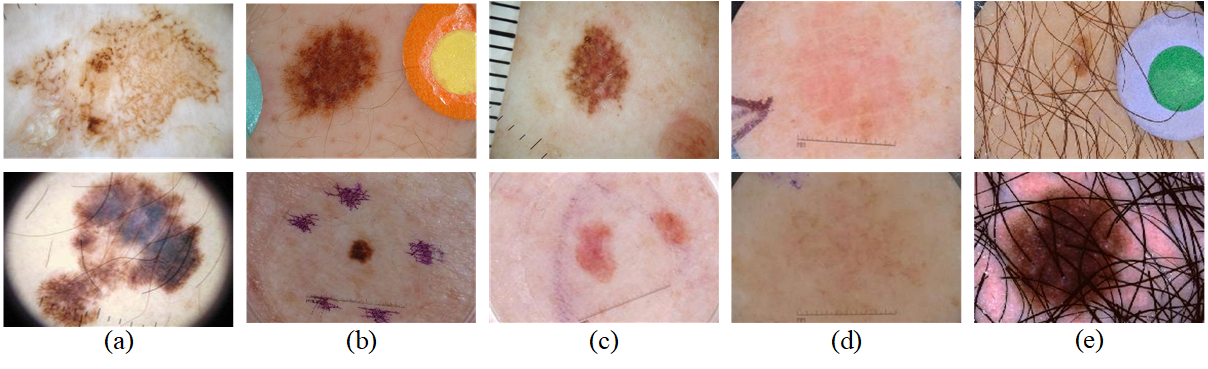}
    \caption{Examples of challenging skin-lesion dermoscopy images: (a) variation in appearance, (b) presence of artifacts, (c) multiple lesions, (d) low contrast, (e) presence of hair.}
    \label{ISICSamples}
\end{figure*}

The need for contextual information to improve segmentation has elevated the prominence of attention mechanisms in deep learning \cite{BASAK2022108673, WANG2022108636}. The attention U-Net model \cite{oktay2018} selects attributes by assigning weights to distinct channels. This enables the model to effectively handle variations in the shapes and sizes of different organs. Zhang \textit{et al.~}\cite{zhang2019} introduced a new method to segment skin lesions by integrating deep residual neural networks with attention mechanisms, with outstanding results. Dual attention methods i.e., CBAM \cite{woo2018} enhance feature representations and capture inter-feature interactions by integrating channel and spatial attention mechanisms. However, current methods are still limited in producing accurate segmentations in demanding scenarios (Fig.\ref{ISICSamples}). Most attention mechanisms focus on one single aspect and fail to fully utilise the diversity of scenarios \cite{farooq2024lssf,iqbal2025tbconvl,khan2024lmbf}. Furthermore, the potential synergies between different attention mechanisms have not been thoroughly explored and harnessed to tackle the difficulties associated with skin lesion segmentation. Furthermore, the integration of multiscale information into attention mechanisms has been limited.

This paper presents a novel approach to skin lesion segmentation, using a hybrid architecture that integrates the capabilities of an EfficientNet-based U-Net with a transformer attention (TA) mechanism and focal modulation (FM). The resulting model, TAFM-Net, harnesses the robust spatial representation features of the EfficientNetV2B1 backbone model in the encoder. The core of our proposed method is a self-aware attention mechanism composed of a TA block and a global spatial attention block that extracts significant regions from the encoder. This serves to refine the subsequent segmentation stages. The decoder subnetwork, comprising seven upsampling blocks, integrates bilinear interpolation and separable convolutional layers for each block's input. In parallel, interconnections are used between each encoder stage and its corresponding decoder counterpart, as well as dense interconnections between decoder blocks. This architectural enhancement significantly improves the preservation of features during the upsampling phase. To further emphasise critical image features, segmentation performance, and interpretability, FM is seamlessly integrated into encoder-decoder skip connections. Moreover, we introduce a fusion loss mechanism, improving the alignment between input and ground-truth images. This enriched loss function guides model training with finer image details, optimising segmentation accuracy.


\section{Related Works}

Deep learning, in particular the use of deep CNNs, is nowadays the most widely used method in skin lesion segmentation tasks, as it allows the establishment of an end-to-end trainable system without manual feature extraction. Encoder-decoder architectures such as U-Net \cite{ronneberger2015} are the most popular. Many methods have customised the structure of U-Net to improve its performance. Examples include SEACU-Net \cite{jiang2022seacu} and BCDU-Net \cite{azad2019bi}, which incorporate attentive or bidirectional ConvLSTMs \cite{song2018pyramid} to enhance their capabilities for tasks involving spatiotemporal data, offering improved segmentation performance, better long-range context understanding, and increased robustness to variable-length sequences. An example of a nested architecture is UNet++ \cite{Zhou2018}, which uses a series of nested and dense skip paths to connect the encoder and decoder. Attention ResUNet \cite{Maji2022} uses a convolutional block as an attention gate to filter the intermediate output of each encoder stage and then concatenates it with the corresponding decoder output. Attention Swin UNet \cite{Aghdam2022} replaces CNN blocks in U-Net with an Attention Swin Transformer module \cite{cao2023swin} to obtain local and global representations. For better discrimination, SEACU-Net \cite{jiang2022seacu} uses dense convolution blocks. Following each convolution operation, a channel and a spatial squeeze \& excitation layer selectively enhances important image information and suppresses less important information across different feature channels during each encoding and decoding phase.

Attention mechanisms help a model focus on critical regions or channels \cite{FIAZ2024110812}. In general, there are two ways to apply them. The first is to place the attention mechanism between the encoder and the decoder \cite{Chen2021}. The second is to place it before the concatenation operation between the intermediate output of each encoder stage and its corresponding decoder output. Both aim to filter out nonrelevant information from the feature maps produced by the encoder \cite{HUANG2024110375}. 

Transformers were first used in natural language processing (NLP), have demonstrated superior performance by effectively capturing the context of textual input, and are also increasingly used for computer vision applications. They employ multiple parallel self-attention mechanisms to capture long-range dependencies. TC-Net \cite{Dong2022}, which is a transformer and CNN fusion network architecture, is designed to more accurately combine local and global information on characteristics. The Multiscale Context Transformer (MCT) is applied in SLT-Net \cite{Feng2022} as a skip connection to achieve interaction of information between skip connections across levels of channel dimension. 

Incorporating attention mechanisms and transformers has made substantial contributions to the segmentation of skin lesions in prior works \cite{YUAN2023109228, GUO2024110491}. However, it is critical to acknowledge the limitations of these works. Although existing methods incorporate attention mechanisms, they struggle to effectively integrate spatial and channel information, which limits their performance. In addition, transformer-based models applied in this context have typically emphasised long-range dependencies, but may not fully exploit the finer-grained details necessary for accurate skin lesion segmentation. Our approach overcomes these limitations by providing a holistic fusion of attention mechanisms, transformers, and focal modulation, allowing a more comprehensive representation of skin lesions, which is necessary to achieve superior segmentation results.

\section{Proposed Method}

\begin{figure}[!t]
    \centering
    \includegraphics[width=\textwidth]{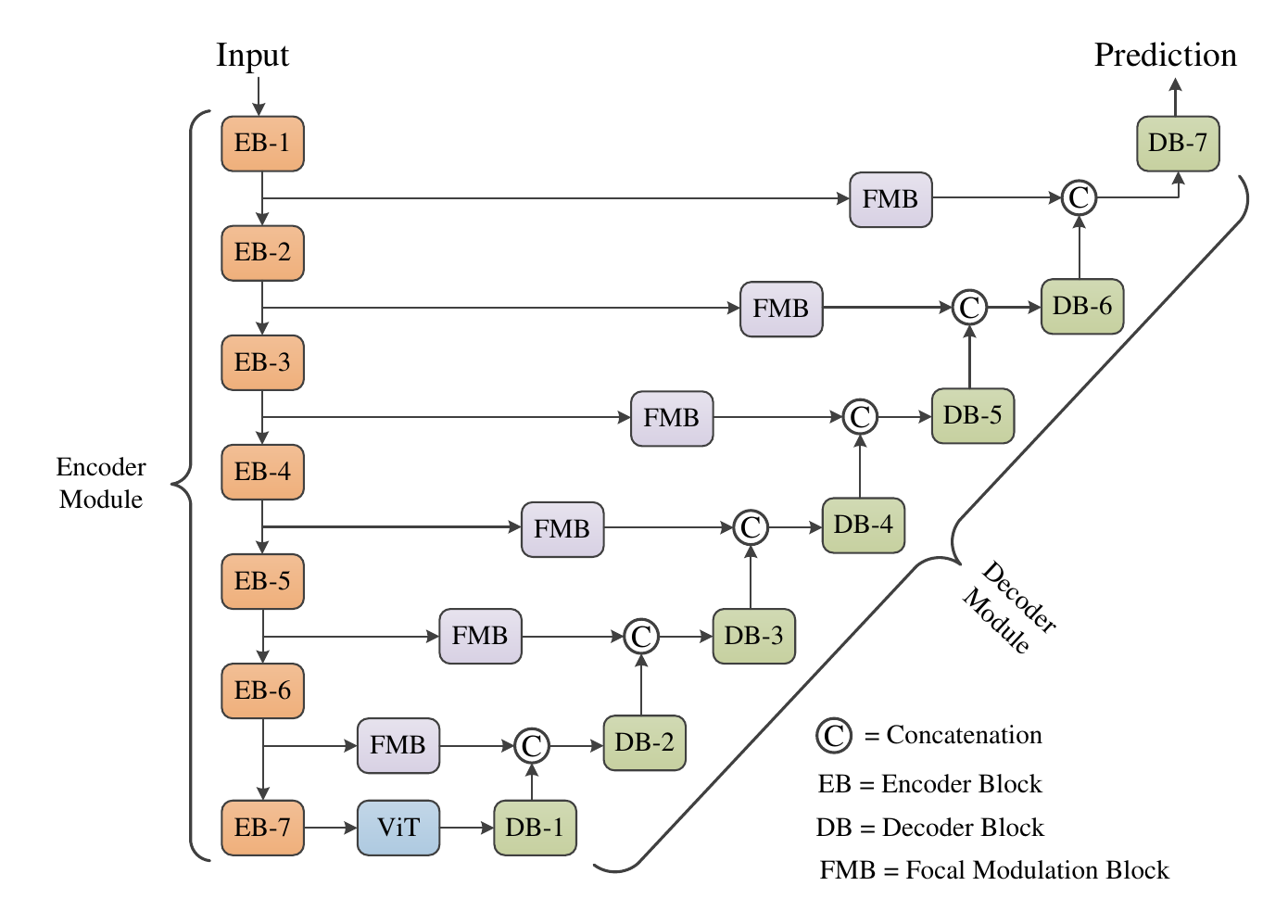}
    \caption{Design of the proposed transformer attention focal modulation network (TAFM-Net).}
    \label{fig:my_label}
\end{figure}

The proposed TAFM-Net (Fig.~\ref{fig:my_label}) is modularised into four components, which are the encoder, the vision transformer (ViT), the focal modulation blocks, and the decoder module. Here we present the details of each component as well as the loss function used to train the network.

\subsection{Encoder Module}
For the encoder of our model, we employ EfficientNetV2 \cite{Tan2021} due to its leading performance in ImageNet. To limit computational cost, EfficientNetV2B1 is selected as the backbone model. Skip connections are built between each encoder stage and the corresponding decoder block, which has the same shape as the feature map. The Sigmoid operation is used in the final segmentation layer, which generates a predicted mask of size $256\times 256\times 1$.

\subsection{Vision Transformer Module}
A self-aware attention module \cite{Chen2021} is used to improve the model's ability to aggregate contextual information. The module consists of two independent blocks, namely a transformer self-attention (TSA) block to capture contextual information from relative positions, infusing positional information through the concatenation of input and positional embeddings, and a global spatial attention (GSA) block to enhance the local contextual information from a broader view through aggregation with global information.


\begin{figure}[!t]
    \centering
    \includegraphics[width=1\textwidth]{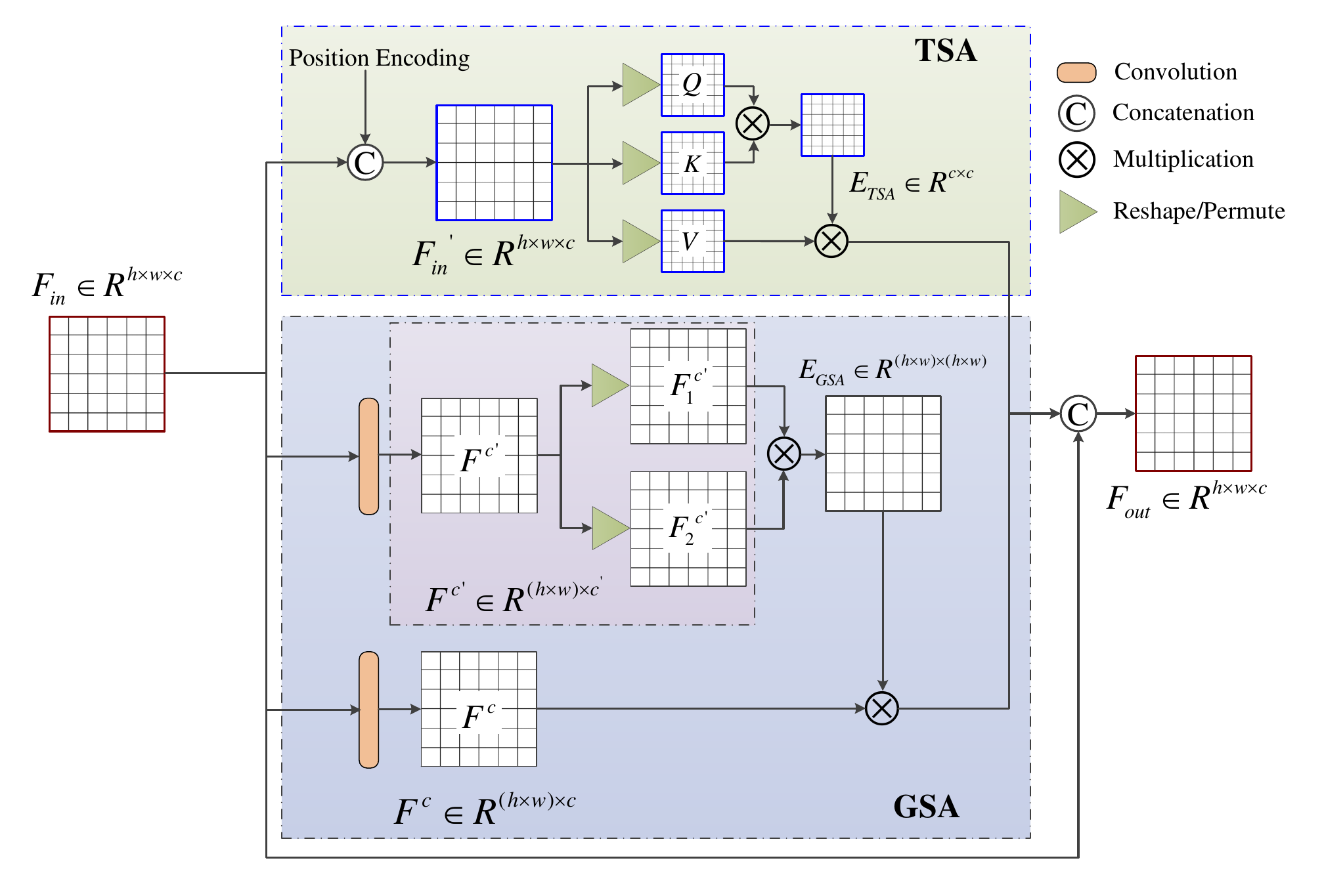}
    \caption{Design of the self-aware attention module. Top: The transformer self-attention (TSA) block. Bottom: The global spatial attention (GSA) block. $F_\text{in}\in R^{8\times 8 \times 1280}$ is the output of the encoder block and $F_\text{out}\in R^{8\times 8\times 1280}$ is the input of the decoder block.}
    \label{fig:SelfAttention}
\end{figure}

\begin{figure}[!t]
    \centering
    \includegraphics[width=1\textwidth]{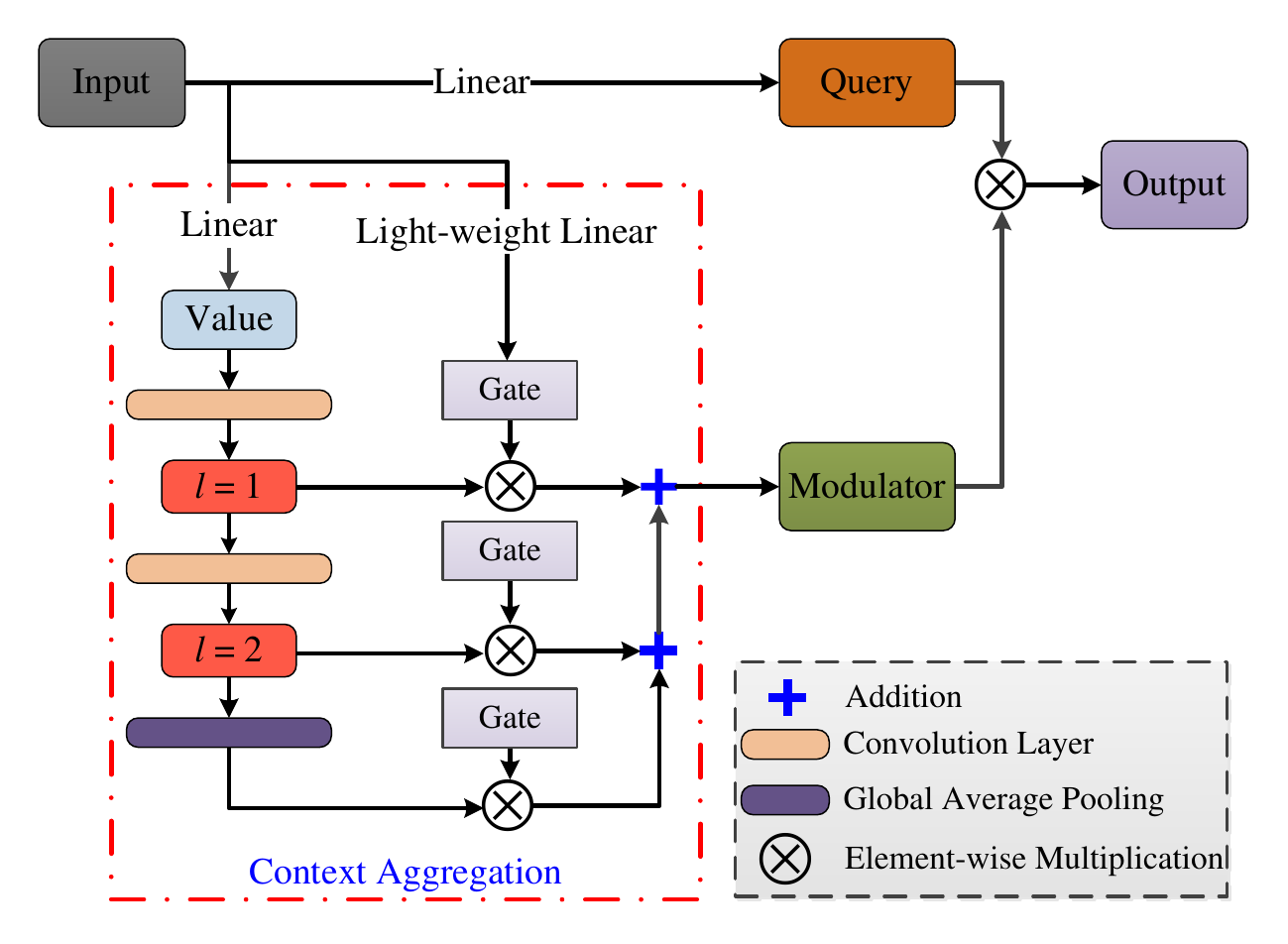}
    \caption{Design of the focal modulation block.}
  \label{fig:focal_mod}
\end{figure}

\subsubsection{Transformer Self-Attention}
As multihead attention can learn self-correlation but cannot learn spatial information, a commonly used approach is to pass the input feature map $F_\text{in}$ through a position encoding and then input the resulting $F'_\text{in}$ into the multihead attention block (Fig.~\ref{fig:SelfAttention}). The map $F'_\text{in}$ is then embedded in three matrices:
\begin{align}
    Q = W_Q\cdot F'_\text{in} & \quad \in R^{(h \times w) \times c} \\
    K = W_K\cdot F'_\text{in} & \quad \in R^{c \times (h \times w)} \\
    V = W_V\cdot F'_\text{in} & \quad \in R^{c \times (h \times w)}
\end{align}
where $W_Q, W_K, W_V$ are three embedding matrices for different linear projections. The scaled dot product of $Q$ and $K$ with Softmax normalisation gives $E_\text{TSA}\in R^{c\times c}$, which represents the similarity between channels in $Q$ and others. By multiplying the value matrix $V$ by the contextual attention map $E_\text{TSA}$, the attention-weighted aggregation values are obtained.
This allows the formulation of the multihead attention mechanism as:
\begin{equation}
    A_\text{TSA}(Q, K, V) = \text{Softmax}\!\left(\frac{QK}{\sqrt{d_k}}\right)V
\end{equation}
and $A_\text{TSA} \in R^{c\times (h\times w)}$ is reshaped to $R^{h\times w \times c}$, the same as the input shape.

\subsubsection{Global Spatial Attention}
Global spatial attention is employed to capture information on global position dependencies. The input feature map $F_\text{in}\in R^{h\times w\times c}$ is embedded in $F^c\in R^{h\times w\times c}$ and $F^{c'}\in R^{h\times w\times c'}$ where $c' = c/2$. The latter is reshaped to $F^{c'}_1\in R^{(h\times w)\times c'}$ and $F^{c'}_2\in R^{c'\times (h\times w)}$, the scaled dot product of which then passes through a Softmax normalisation layer, and the resulting output map $E_\text{GSA}\in R^{(h\times w)\times (h\times w)}$ indicates the spatial similarity (correlation) between any two positions. The multihead attention mechanism can then be defined as:
\begin{equation}
    A_\text{GSA} = \text{Softmax}\!\left(F^{c'}_1\cdot F^{c'}_2\right)F^c
\end{equation}
and the final output $F_\text{out}\in R^{h \times w \times c}$ of the self-aware attention module is the concatenation of $A_\text{TSA}$, $A_\text{GSA}$, and $F_\text{in}$.

\subsection{Focal Modulation Blocks}
The focal modulation (FM) \cite{naderi2022focal} blocks offer several advantages for medical image segmentation, making them a valuable addition to CNN. An FM block computes a global feature map from the input image, which then modulates the output of the local convolution (Fig.~\ref{fig:focal_mod}). That is, the global feature map controls the attention of the local convolution. Here, a depth-wise separable convolution is used to generate the global feature map, where each input channel is convolved with a single kernel, reducing computational overhead while computing the global feature map. Local convolution is a standard operation that is responsible for extracting features from a small region of the input image. Its subsequently modulation by the global feature map ensures that the local convolution's output is consistent with the information present in the global feature map.

The context aggregation block aggregates information from neighbouring regions of a feature map. It computes a weighted sum of the neighbouring feature map values, where the weights are determined by a gating function. This function decides the attention given to each neighbour, potentially emphasising neighbours closer to the centre of the feature map. The weighted sum of the neighbours is then added to the original feature map, resulting in a new feature map that has been enriched with the local context. This new feature map can be used by the FM block to capture long-range dependencies between different regions of the input image. In this way, the context aggregation block facilitates the incorporation of context-aware information, allowing the FM block to learn and leverage relevant long-range relationships in the segmentation process.


\subsection{Decoder Module}
\begin{figure}[!t]
    \centering
    \includegraphics[width=1\textwidth]{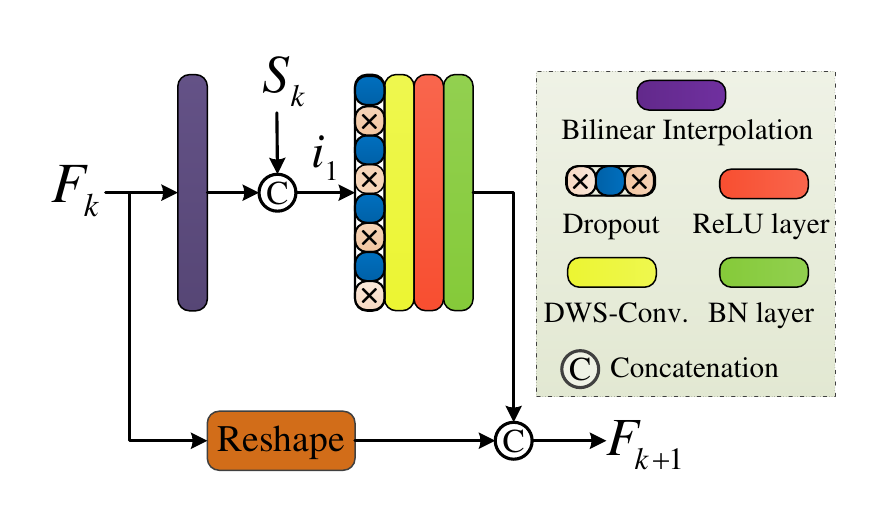}
    \caption{Design of the decoder block.}
    \label{fig:decoder}
\end{figure}

The decoder module consists of seven sequential upsampling blocks. In each upsampling block (Fig.~\ref{fig:decoder}), the input feature map $F_k$ goes through a bilinear interpolation layer (B) and is then concatenated with the skip connection $S_k$ from the corresponding encoder block, generating a feature map with the same shape that is passed to a convolutional layer for feature fusion. In our case, we employ separable convolution to reduce the number of parameters.
The output $F_{k+1}$ of the decoder block is computed from the fused feature map by applying dropout (D) with $0.5$ probability followed by $3\times 3$ depthwise separable convolution (DWSC), rectified linear unit (ReLU) activation, and batch normalization (BN),  and concatenating with the input $F_{k}$ reshaped to match the size:
\begin{equation}
    F_{k+1}=\text{BN}\!\left(\text{ReLU}\!\left(\text{DWSC}(\text{D}(\text{B}(F_{k})\copyright S_{k}))\right)\right)\copyright\ \text{Reshape}(F_{k})
    \label{Eq:8}
\end{equation}
where the latter concatenation is essentially a residual connection. An alternative approach to preserving low-level representations is through the use of dense connections, which incorporate the upsampling feature from the previous encoder block as input for the current block and utilise the output feature map as input for all subsequent blocks. We conducted experiments on both residual and dense connections. To convert the predicted map from the decoder module to a binary segmentation mask, a threshold is used that maximises the F1 score.

\subsection{Loss Function}
For training of our proposed network, we use a loss function consisting of regional and boundary losses. We briefly introduce the most widely used loss functions for skin lesion segmentation and the different fused losses we considered in our experiments. In the following, $P$ and $G$ denote the prediction of the model and the segmentation of the ground truth, respectively, and $p^{c}_{i}$ is the prediction of the model that the pixel $i$ belongs to the class $c$, while $g^{c}_{i}$ is the corresponding ground truth label. Where needed, we use a small value $\epsilon=1e^{-5}$ to prevent division by zero.

\subsubsection{Binary Cross Entropy Loss}
The concept of cross entropy is utilised as a quantitative measure to differentiate between two probability distributions. The expression for the binary cross-entropy (BCE) loss within the domain of binary segmentation is as follows:
\begin{equation}
L_\text{BCE} = -\sum_{i=1}^{N}\left[g^{c}_{i}\log p^{c}_{i} + (1-g^{c}_{i})\log(1-p^{c}_{i})\right]
\end{equation}

\subsubsection{Dice Loss}
The Dice similarity coefficient (DSC) is a metric used to evaluate the extent of concurrence between the segmentation and the ground truth. The Dice score, which is calculated specifically for binary segmentation tasks, is as follows:
\begin{equation}
\text{DSC}_c = \frac{2\vert G\cap P\vert}{\vert G\vert + \vert P\vert}
= \frac{\left[\sum_{i=1}^Np^{c}_{i}g^{c}_{i}\right]+\epsilon}{\left[\sum^N_{i=1}p^{c}_{i}+g^{c}_{i}\right]+\epsilon}
\end{equation}
Therefore, in order to minimise the discrepancy between the ground truth and segmented region, the Dice loss $L_\text{DSC}$ is formulated, expressed as:
\begin{equation}
L_\text{DSC} = \sum_c (1-\text{DSC}_c)
\end{equation}
An inherent limitation of the Dice loss algorithm is its equivalent emphasis on false positive (FP) and false negative (FN) pixels. Specification maps with high specificity but low recall are the norm when employing this method, especially on datasets with imbalanced classes \cite{Abraham2018}.

\subsubsection{Jaccard Loss} 
The Jaccard similarity coefficient (JSC) is a metric used to evaluate the degree of similarity (overlap) between the ground truth and segmentation sets. It is computed by dividing the intersection of the two sets by the union:
\begin{align}
    \text{JSC}_c &= \frac{\vert G\cap P\vert}{\vert G\cup P\vert} = \frac{\vert G\cap P\vert}{\vert G\vert + \vert P\vert - \vert G \cap P\vert}\notag\\
    &= \frac{\left[\sum_{i=1}^Np^{c}_{i}g^{c}_{i}\right]+\epsilon}{\left[\sum^N_{i=1}p^{c}_{i}+g^{c}_{i}-p^{c}_{i}g^{c}_{i}\right]+\epsilon}.
\end{align}
The Jaccard loss $L_\text{JSC}$ is then calculated as:
\begin{equation}
    L_\text{JSC} = \sum_c (1-\text{JSC}_c).
\end{equation}

\subsubsection{Focal Tversky Loss}
The focal Tversky (FT) loss \cite{Abraham2018} is designed to weight the FN and FP pixels so that the model can learn the details of the minority class in the case of an imbalanced dataset. In the binary segmentation case, the mathematical representation of the Tversky index (TI) is:
\begin{equation}
    \text{TI}_c = \frac{\left[\sum^N_{i=1}p^{c}_{i}g^{c}_{i}\right]+\epsilon}{\left[\sum^N_{i=1}p^{c}_{i}g^{c}_{i} + \alpha\sum^N_{i=1}p^{\Bar{c}}_{i}g^{c}_{i} + \beta\sum_{i=1}^Np^{c}_{i}g^{\Bar{c}}_{i}\right]+\epsilon}.
\end{equation}
This index can be used to define the following loss function:
\begin{equation}
    L_\text{TI} = \sum_c (1-\text{TI}_c).
\end{equation}
From this, the FT loss is obtained by introducing the hyperparameter $\gamma$, which can be tuned to allow for a greater emphasis on the challenging classes that have a low likelihood of being discovered:
\begin{equation}
    L_\text{FT} = \sum_c (1-\text{TI}_c)^{\frac{1}{\gamma}}.
\end{equation}
When $\gamma=1$, the loss $L_\text{FT}$ simplifies to $L_\text{TI}$. In our work, we used $\alpha = 0.3, \beta = 0.7, \gamma=1$ based on the original paper \cite{Abraham2018}.


\subsubsection{Boundary Loss} 
The concept of boundary loss \cite{Kervadec2021} was initially introduced as a means of improving the representation of the contours of the object. The fundamental goal of this loss is to minimise the non-symmetric $L_2$ distance between the contour of the real object and the contour of the segmented object. Denoting the boundary of the ground-truth object region $G$ as $\partial G$, and of the model-predicted object region $P$ as $\partial P$, the changes between two boundaries can be formulated as
\begin{equation}
\text{Dist}(\partial P, \partial G) = \int_{\partial G} \Vert y_{\partial P}(p) - p\Vert dp
\end{equation}
where $p$ is a point located on $\partial G$, and $y_{\partial P}$ indicates its closest point on $\partial P$. To circumvent local differential computations related to contour points, this equation can be approximated as a regional integral, yielding
\begin{equation}
\text{Dist}(\partial P, \partial G) \approx 2\!\int_{\Delta P} D_{G}(q) dq
\end{equation}
where $\Delta P$ represents the area between the ground-truth and predicted contours, $D_G: \Omega \rightarrow \mathbb{R}^+$ denotes the distance map with respect to $G$, and $D_G(q)$ is the distance between a point $q\in \Omega$ and its nearest point $z_{\partial G}(q)$ on $\partial G$. It has been shown \cite{Kervadec2021} that the latter equation can be rewritten as
\begin{equation}
    \text{Dist}(\partial P, \partial G) = 2\!\int_{\Omega} \big(\phi_G(q)\xi(q) - \phi_G(q)\zeta(q)\big) dq
    \label{eq:dist}
\end{equation}
where $\xi, \zeta: \Omega\rightarrow\{0, 1\}$ are the indicator functions for the regions $P$ and $G$, respectively, and $\phi_G:\Omega \rightarrow \mathbb{R}$ is the level-set function of $\partial G$ such that $\phi_G(q)=-D_G(q)$ if $q\in G$ and $\phi_G(q)=D_G(q)$ otherwise. Replacing the binary value $\xi$ by the softmax probability output $\xi_\theta$ of the network having parameters $\theta$, and ignoring the second term in (\ref{eq:dist}) as it is independent of the network parameters $\theta$, we can formulate the boundary loss as
\begin{equation}
    L_\text{B} = \int_\Omega \phi_G(q)\xi_{\theta}(q) dq.
\end{equation}

\subsubsection{Fused Losses}
In the experiments, to potentially achieve a better regional segmentation, we combined $L_\text{JSC}$, $L_\text{FT}$, and $L_\text{DSC}$ with $L_\text{BCE}$, respectively, yielding
\begin{align}
    L_1 &= L_\text{BCE} + L_\text{JSC}, \\
    L_2 &= L_\text{BCE} + L_\text{FT}, \\
    L_3 &= L_\text{BCE} + L_\text{DSC}.
\end{align}
Furthermore, we fused each of these losses with $L_\text{B}$, yielding
\begin{align}
    L_4 &= \alpha L_1 + (1-\alpha)L_\text{B}, \\
    L_5 &= \alpha L_2 + (1-\alpha)L_\text{B}, \\
    L_6 &= \alpha L_3 + (1-\alpha)L_\text{B},
\end{align}
where $\alpha$ represents the fusion weight of the loss. To make these fused losses rely more on $L_{1,2,3}$ in the early stages of training to accelerate model convergence, but increase the guidance of $L_\text{B}$ as training progresses, we initially set $\alpha=1$ and then decreased $\alpha$ with a constant $\Delta\alpha = 0.005$ at the end of each training epoch, so that we could control the contribution of both loss components dynamically. The floating range of $\alpha$ is between $0.01$ and $1$, so both components are considered throughout the training process.

\section{Experiments and Results}
The proposed TAFM-Net was evaluated on several popular public datasets for skin lesion segmentation and compared with many other state-of-the-art (SOTA) methods. We summarise the datasets, performance measures, and implementation details of our method as used in the experiments and then present an ablation study and performance comparisons with other methods.

\subsection{Skin Lesion Segmentation Datasets}
For the performance evaluation, we used four publicly accessible skin lesion datasets. Three were obtained from the archive of the International Skin Imaging Collaboration (ISIC), namely ISIC2016\footnote{\tiny\url{https://challenge.isic-archive.com/data/\#2016}}\label{f1}, ISIC2017\footnote{\tiny\url{https://challenge.isic-archive.com/data/\#2017}},
ISIC2018\footnote{\tiny\url{https://challenge.isic-archive.com/data/\#2018}}.Additionally, we included the PH2\footnote{\tiny\url{https://datasetninja.com/ph2}} dataset (see Table~\ref{datasets} for details).





\begin{table}[!t]
  \centering
  \caption{Summary of datasets utilized for evaluating TAFM-Net for skin lesion segmentation. For each dataset, the table lists the number of images in the training, validation, and test sets, along with the $x\times y$ size of the images in pixels.}
 \begin{tabular}{lcccc}
\toprule
\multicolumn{1}{l}{\multirow{2}{*}{\textbf{Dataset}}}&\multicolumn{3}{c}{\textbf{Number of Images}}&\multicolumn{1}{c}{\multirow{2}{*}{\textbf{Resolution}}}\\
\cmidrule{2-4}
& \textbf{Train} & \textbf{Validation} & \textbf{Test} & \\
\toprule
ISIC2016 & 900 & NA & 379 & 679$\times$453--6,748$\times$4,499\\
ISIC2017 & 2,000 & NA & 600 & 679$\times$453--6,748$\times$4,499\\
ISIC2018 & 2,594& 100& 1,000& 679$\times$453--6,748$\times$4,499\\
PH2 & 200 & NA & NA & 768$\times$560 \\
\toprule
\end{tabular}
  \label{datasets}%
\end{table}%

\subsection{Performance Measures}
Quantification of performance was done using five evaluation metrics: accuracy, sensitivity, specificity, Jaccard index, and Dice coefficient. These metrics are computed from the number of true positive (TP), true negative (TN), false positive (FP), and false negative (FN) pixels in the model's predictions:
\begin{equation}
\mathrm{Accuracy\ (A) = \frac{TP+TN}{TP+TN+FP+FN}},
\end{equation}
\begin{equation}
\mathrm{Sensitivity\ (Sn) = \frac{TP}{TP+FN}},
\end{equation}
\begin{equation}
\mathrm{Specificity\ (Sp) = \frac{TN}{TN+FP}},
\end{equation}
\begin{equation}
\mathrm{Jaccard\ (J) = \frac{TP}{TP+FP+FN}},
\end{equation}
\begin{equation}
\mathrm{Dice\ (D) = \frac{2TP}{2TP+FP+FN}}.
\end{equation}
All metrics range from 0 (worst performance) to 1 (best performance).

\subsection{Implementation Details}
All training images were initially reshaped to $256\times 256$ and then fed to the TAFM-Net. The Adam method was used as an optimiser, with initial decay rates $\beta_1 = 0.9$ and $\beta_2 = 0.999$ to estimate, respectively, the first and second moments of the gradient at the end of each epoch. This is in line with a previous study \cite{Mirikharaji2022} showing that these decay rates are used most frequently in published work on skin lesion analysis. Similarly, the initial learning rate was set to 0.001. During training, early stop monitoring started from the 10th epoch, and training was terminated when the monitored metric did not improve for 9 epochs.
All experiments were performed in Google Colab on the Keras framework with Python 3.9.

\subsection{Ablation Experiments}
Several ablation experiments were conducted to evaluate different aspects of TAFM-Net before comparing the network to other SOTA methods. We evaluated the impact of different loss functions, different network components, and different training versus test datasets on the performance of the network.

\begin{table}[!t]
\centering
\caption{Results of an ablation experiment of different losses with TAFM-Net on the ISIC 2016 dataset. Numbers are means of a 3-fold cross-validation. Best score per metric in \textbf{bold}.}
\begin{tabular}{cccccc}
\toprule     
\multicolumn{1}{c}{\multirow{2}{*}{\textbf{Loss Function}}} &
\multicolumn{5}{c}{\textbf{Performance Measures (\%)}} \\
\cmidrule{2-6}
& $\mathbf{A}$ & $\mathbf{Sn}$ & $\mathbf{Sp}$ & $\mathbf{J}$ & $\mathbf{D}$ \\
\toprule
$L_\text{BCE}$ & 98.36 & 96.20 & 97.82 & 93.14 & 96.35 \\
$L_\text{DSC}$ & 95.33 & 91.60 & 98.09 & 81.49 & 85.63 \\
$L_\text{JSC}$ & 94.63 & 92.34 & \textbf{98.41} & 78.64 & 82.40 \\
$L_\text{FT}$ & 95.85 & 94.37 & 97.09 & 81.72 & 85.63 \\
$L_\text{B}$ & 94.03 & 94.66 & 96.76 & 75.38 & 79.18 \\
$L_1$ & 98.36 & 96.32 & 97.60 & 93.25 & 96.35 \\
$L_2$ & 98.38 & 96.32 & 97.77 & 93.17 & 96.34 \\
$L_3$ & 98.38 & 96.26 & 97.85 & 93.32 & 96.45 \\
$L_4$ & \textbf{98.44} & \textbf{96.65} & 97.66 & \textbf{93.64} & \textbf{96.66} \\
$L_5$ & 98.40 & 96.52 & 97.79 & 93.46 & 96.56 \\
$L_6$ & 98.42 & 96.63 & 97.71 & 93.56 & 96.58 \\
\bottomrule
\end{tabular}
\label{tab:Ablation_loss}%
\end{table}

\begin{figure}[!t]
    \centering
    \vspace{1em}
    \includegraphics[width=\textwidth]{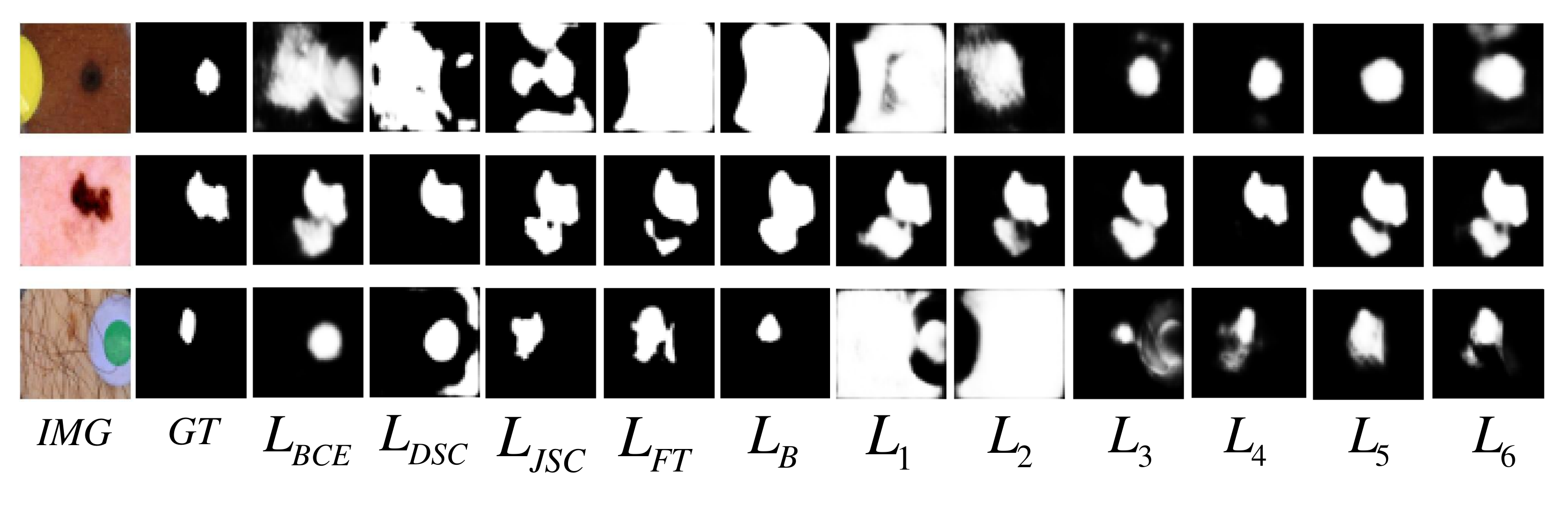}
    \caption{Visual evaluation of TAFM-Net for the different loss functions on three challenging example images from the ISIC 2016 dataset. IMG = example image (left column). GT = corresponding ground truth (second column). The first example (top row) is a case of a relatively dark image. The second example (middle row) shows a confusing object. The third example (bottom row) contains many hairs that may impact the prediction.}
    \label{LossVis}
\end{figure}

\subsubsection{Impact of Loss Function}
The results of the ablation experiment with different loss functions (Table \ref{tab:Ablation_loss}) show that $L_4$ achieved the best overall performance in terms of most metrics. We also found that replacing $L_1 \rightarrow L_4$, $L_2\rightarrow L_5$, and $L_3 \rightarrow L_6$ consistently yielded small improvements in most metrics. This suggests that boundary information contributes to guiding the segmentation. This is confirmed by visual inspection (Fig.~\ref{LossVis}), from which we observe that the BCE loss alone struggles to identify the lesion. The boundary loss shows better performance in these complicated cases. The fused losses without using boundary information ($L_1, L_2, L_3$) did not demonstrate improvement, while the inclusion of the boundary loss ($L_4, L_5, L_6$) yielded more accurate segmentation performance.

\subsubsection{Impact of Network Components}
\begin{figure}[!t]
    \centering
    \resizebox{\textwidth}{!}{%
    \begin{tabular}{c}
        \includegraphics[width=\textwidth]{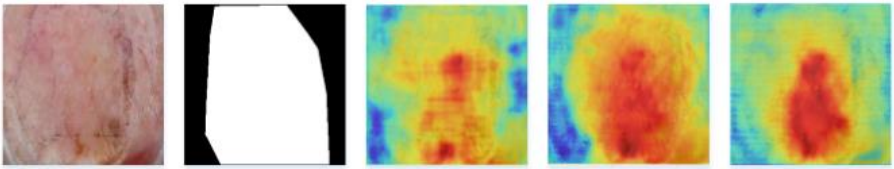} \\
        \includegraphics[width=\textwidth]{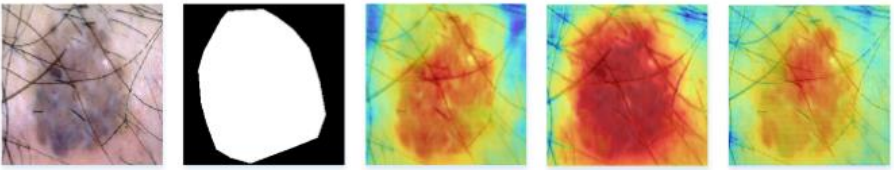} \\ 
        \includegraphics[width=\textwidth]{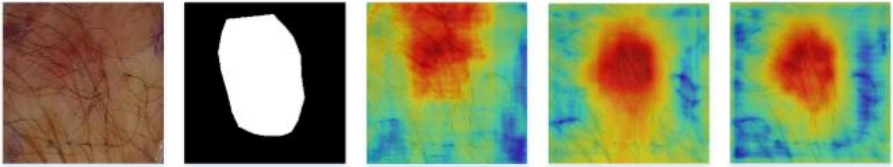} \\       
        \includegraphics[width=\textwidth]{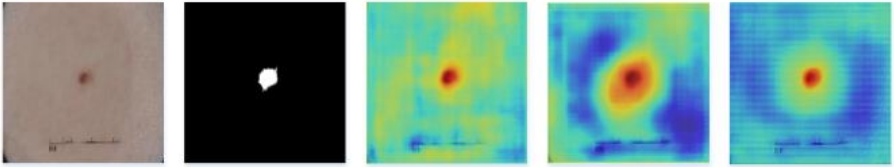} \\ 
    \end{tabular}
    }
    \caption{Heatmaps showing the impact of the different network components of TAFM-Net. The three rows show three different examples from the ISIC 2016 dataset. Columns from left to right: input RGB image, corresponding ground-truth image, heatmap of the baseline model, heatmap of the baseline model with transformer self-aware attention in the bottleneck layer, and heatmap of the latter model with focal modulation in the skip connections.}
    \label{fig:heatmaps}
\end{figure}

Next, we conducted an ablation experiment with different components of the TAFM-Net network. The baseline consisted of an EfficientNetV2B1 encoder and a decoder subnetwork with residual connections. To ensure consistency, we used $L_4$ as the loss function throughout this experiment. The results (Table \ref{tab:Ablation_components}) indicate that incorporating FM blocks into the skip connections of the proposed TAFM-Net, along with the transformer in the encoder-decoder bottleneck layer, leads to superior performance across all performance metrics. We also used Grad-CAM \cite{selvaraju2017grad} to visualise attention heatmaps, allowing direct observation of specific feature areas that various components of the network emphasise. The maps (Fig.~\ref{fig:heatmaps}) indicate that the proposed approach works better when we combine transformer self-aware attention in the bottleneck layer with FM in the skip connections.

\begin{table}[!t]
  \centering
  \caption{Results of an ablation experiment of different network components of TAFM-Net on the ISIC 2016 dataset. Numbers are means of a 3-fold cross-validation. Best score per metric in \textbf{bold}.}
    \begin{tabular}{ccccccc} 
    \toprule
    \multirow{2}[4]{*}{\textbf{Transformer}} & \multirow{2}[4]{*}{\textbf{Skip Connection}} & \multicolumn{5}{c}{\textbf{Performance (\%)}} \\
\cmidrule{3-7} & & $\mathbf{A}$ & $\mathbf{Sn}$ & $\mathbf{Sp}$ & $\mathbf{J}$ & $\mathbf{D}$ \\
    \midrule
    $\times$ & Direct & 97.70 & 95.20 & 97.27 & 91.40 & 95.30 \\
    $\checkmark$ & Direct & 97.72 & 95.37 & 97.21 & 91.54 & 95.32 \\
    $\checkmark$ &  BiConvLSTM \cite{song2018pyramid} & 97.89 & 95.52 & 96.88 & 91.50 & 95.09 \\
    $\checkmark$ & CBAM \cite{woo2018} & 97.64 & 95.12 & 96.76 & 91.03 & 95.04 \\
    $\checkmark$ & FM \cite{yang2022focal} & \textbf{98.44} & \textbf{96.65} & \textbf{97.66} & \textbf{93.64} & \textbf{96.66} \\
    \bottomrule
    \end{tabular}%
  \label{tab:Ablation_components}%
\end{table}%

\begin{table}[!t]
\centering
\caption{Results of an ablation experiment in which TAFM-Net was trained on each one of the ISIC datasets and tested on the other two. Numbers are means of a 3-fold cross-validation. Best score per metric and per training dataset in \textbf{bold}.}
\begin{tabular}{llccccc}
\toprule
\multirow{2}[4]{*}{\textbf{Training}} & \multirow{2}[4]{*}{\textbf{Testing}} & \multicolumn{5}{c}{\textbf{Performance (\%)}} \\
\cmidrule{3-7} & & $\mathbf{A}$ & $\mathbf{Sn}$ & $\mathbf{Sp}$ & $\mathbf{J}$ & $\mathbf{D}$ \\
\toprule
& ISIC 2016 & \textbf{98.44} & \textbf{96.65} & \textbf{97.66} & \textbf{93.64} & \textbf{96.66} \\
ISIC 2016 & ISIC 2017 & 94.95 & 87.95 & 93.14 & 84.88 & 88.17 \\
& ISIC 2018 & 95.87 & 91.39 & 95.92 & 89.88 & 93.53 \\
\midrule
& ISIC 2016 & 96.44 & 94.65 & 94.66 & 90.64 & 94.66 \\
ISIC 2017 & ISIC 2017 & \textbf{96.95} & \textbf{92.95} & \textbf{95.14} & \textbf{86.88} & \textbf{92.17} \\
& ISIC 2018 & 94.87 & 93.39 & 94.92 & 88.88 & 94.53 \\
\midrule
& ISIC 2016 & 95.44 & 92.65 & 93.66 & 90.64 & 91.66 \\
ISIC 2018 & ISIC 2017 & 91.95 & 89.95 & 93.14 & 83.88 & 90.17 \\
& ISIC 2018 & \textbf{97.87} & \textbf{96.39} & \textbf{97.92} & \textbf{92.88} & \textbf{96.53} \\
\bottomrule
\end{tabular}%
\label{tab:cross_dataset}%
\end{table}%

\subsubsection{Impact of Different Datasets}
We also evaluated the generalisability of TAFM-Net when trained on one dataset and tested on the others. Specifically, we trained the network on each one of the three ISIC datasets separately, and tested the resulting models on the other two. Not surprisingly, the results (Table \ref{tab:cross_dataset}) show that the best performance was achieved on the test data of the corresponding training data of the same dataset. However, the performance on the other datasets was only somewhat lower, and overall the network was able to generalise quite well.





\begin{table}[htbp]
  \centering
  \caption{Performance comparison of the proposed TAFM-Net and existing SOTA methods on the ISIC 2018, ISIC 2017, and ISIC 2018 datasets, respectively.  Best score per metric in \textbf{bold}.}
  \adjustbox {max width=\textwidth}{
    \begin{tabular}{lccccccccccccccccc}
    \toprule
    \multirow{3}[6]{*}{\textbf{Method}} & \multicolumn{17}{c}{\textbf{Performance Measures (\%)}} \\
\cmidrule{2-18}          & \multicolumn{5}{c}{\textbf{ISIC2016}} &       & \multicolumn{5}{c}{\textbf{ISIC2017}} &       & \multicolumn{5}{c}{\textbf{ISIC2018}} \\
\cmidrule{2-6}\cmidrule{8-12}\cmidrule{14-18}          & \textbf{A} & \textbf{Sn} & \textbf{Sp} & \textbf{J} & \textbf{D} &       & \textbf{A} & \textbf{Sn} & \textbf{Sp} & \textbf{J} & \textbf{D} &       & \textbf{A} & \textbf{Sn} & \textbf{Sp} & \textbf{J} & \textbf{D} \\
    \midrule
    ARU-GD \cite{Maji2022} & 94.38 & 89.86 & 94.65 & 85.12 & 90.83 &       & 93.88 & 88.31 & 96.31 & 80.77 & 87.89 &       & 94.23 & 91.42 & 96.81 & 84.55 & 89.16 \\
    AutoSMIM \cite{wang2023autosmim}& 96.42 & -     & -     & 87.05 & 92.73 &       & 93.91 & -     & -     & 77.87 & 85.72 &       & 96.21 & -     & -     & 83.96 & 90.40 \\
    BCDU-Net \cite{azad2019bi} & 91.78 & 78.11 & 96.20 & 83.43 & 80.95 &       & 91.63 & 76.46 & 97.09 & 79.20 & 78.11 &       & 93.70 & 78.50 & 97.20 & 81.10 & 85.10 \\
    CPFNet \cite{9049412} & 95.09 & 92.11 & 95.91 & 83.81 & 90.23 &       &    -   &     -  &   -    &   -    &  -     &       & 94.96 & 89.53 & 96.55 & 79.88 & 87.69 \\
    DAGAN \cite{Lei2020}  & 95.82 & 92.28 & 95.68 & 84.42 & 90.85 &       & 93.26 & 83.63 & \textbf{97.24} & 75.94 & 84.25 &       & 93.24 & 90.72 & 95.88 & 81.13 & 88.07 \\
    Hyper-Fusion Net \cite{bi2022hyper} & 96.64 & 94.22 & 96.45 & 88.17 & -     &       & 95.80 & 92.33 & 96.16 & 83.70 & -     &       &    -   &     -  &    -   &     -  &  \\
    Swin-Unet \cite{cao2023swin} & 96.00 & 92.27 & 95.79 & 87.60 & 88.94 &       & 94.76 & 88.06 & 96.05 & 80.89 & 81.99 &       & 96.83 & 90.10 & 97.16 & 82.79 & 88.98 \\
    U-Net \cite{ronneberger2015}  & 93.31 & 87.28 & 92.88 & 81.38 & 88.24 &       & 93.29 & 84.30 & 93.41 & 75.69 & 84.12 &       & 92.52 & 85.22 & 92.09 & 80.09 & 86.64 \\
    UNet++ \cite{Zhou2018} & 93.88 & 88.78 & 93.52 & 82.81 & 89.19 &       & 93.73 & 87.13 & 94.41 & 78.58 & 86.35 &       & 93.72 & 88.70 & 93.96 & 81.62 & 87.32 \\
    \hline
    Proposed TAFM-Net & \textbf{98.44} & \textbf{96.65} & \textbf{97.66} & \textbf{93.64} & \textbf{96.66} &       & \textbf{96.95} & \textbf{92.95} & 95.14 & \textbf{86.88} & \textbf{92.17} &       & \textbf{99.08} & \textbf{96.90} & \textbf{98.19} & \textbf{93.08} & \textbf{96.85} \\
    
    \bottomrule
    \end{tabular}%
    }
  \label{tab:skin}%
\end{table}%

\subsection{Comparative Performance Evaluation}
Finally, we conducted a comprehensive performance comparison of the proposed TAFM-Net with SOTA methods on multiple benchmark datasets, including ISIC 2018, ISIC 2017, ISIC 2016, and PH2. We note that since many of the comparison methods were not actually available, the performance scores for these methods were taken from the original articles, inasmuch as reported. However, in order to provide visual comparisons, we have reproduced the results of some representative methods for which codes are publicly available, and reported these results in the corresponding tables. All other results reported in the tables are taken from the cited papers.

\subsubsection{Comparison on ISIC Datasets}
\begin{figure*}[!t]
    \centering
    \includegraphics[width=\textwidth]{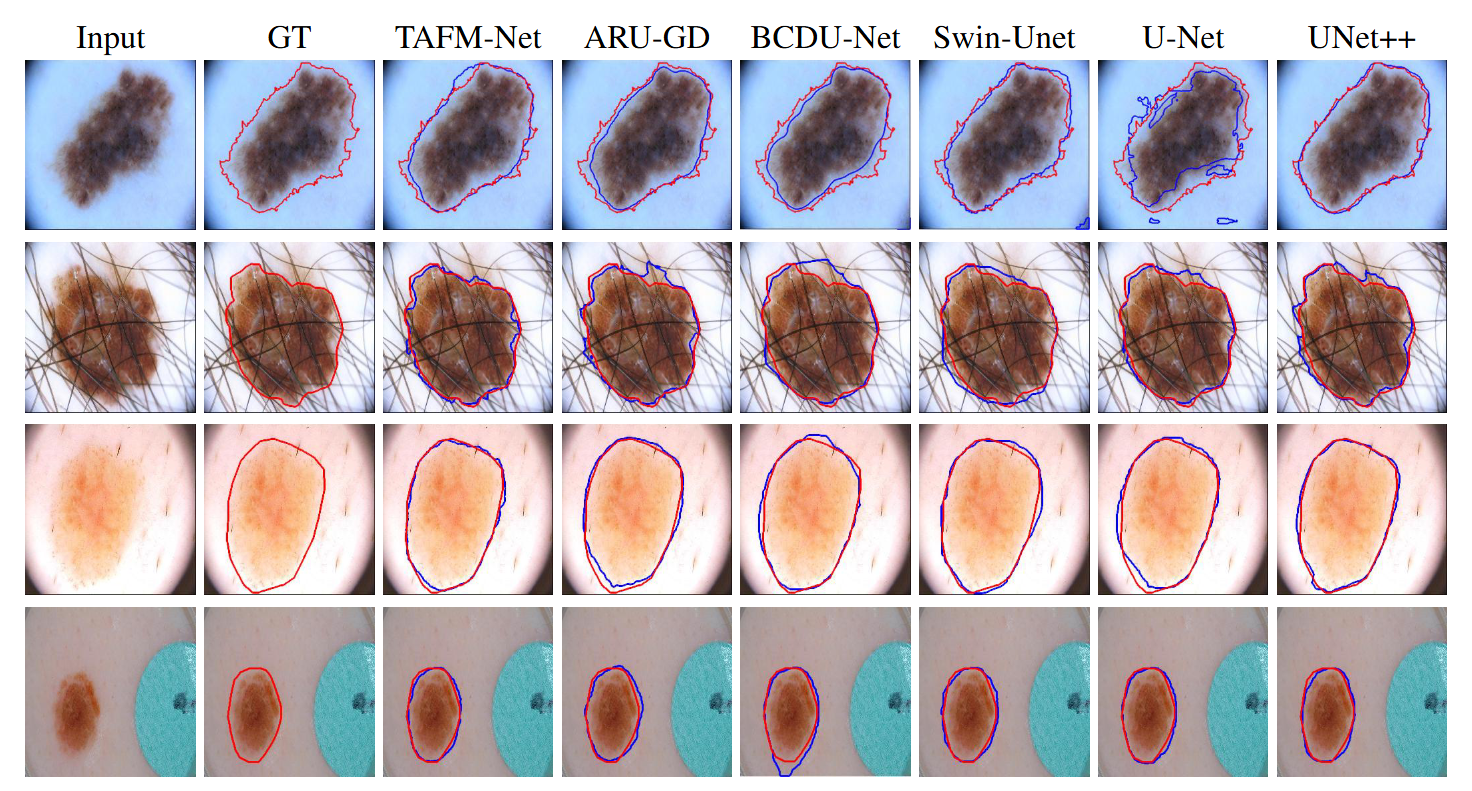}
    \caption{Visual performance comparison of TAFM-Net with other SOTA methods on four example cases from the ISIC 2018 dataset.}
    \label{fig:Vis_ISIC2018}
\end{figure*}

We compared TAFM-Net with the following SOTA methods reported in the literature for that dataset: ARU-GD \cite{Maji2022}, AutoSMIM \cite{wang2023autosmim}, BCDU-Net \cite{azad2019bi}, CPFNet \cite{9049412}, DAGAN \cite{Lei2020}, Hyper-Fusion Net \cite{bi2022hyper}, Swin-Unet \cite{cao2023swin}, U-Net \cite{ronneberger2015}, and UNet++ \cite{Zhou2018}. The results (Table \ref{tab:skin}) show that TAFM-Net consistently outperformed existing methods by a considerable margin, achieving improvements in the Jaccard score of 9.3\%--19.7\% on ISIC2018 dataset. Example visual comparisons on ISIC2018 dataset (Fig.~\ref{fig:Vis_ISIC2018}) with selected methods for which source code was available confirm the superiority of TAFM-Net over other methods, especially in images with obstacles such as occlusions, black backgrounds, hairs and low contrast.

\begin{figure*}[!t]
    \centering
    \includegraphics[width=\textwidth]{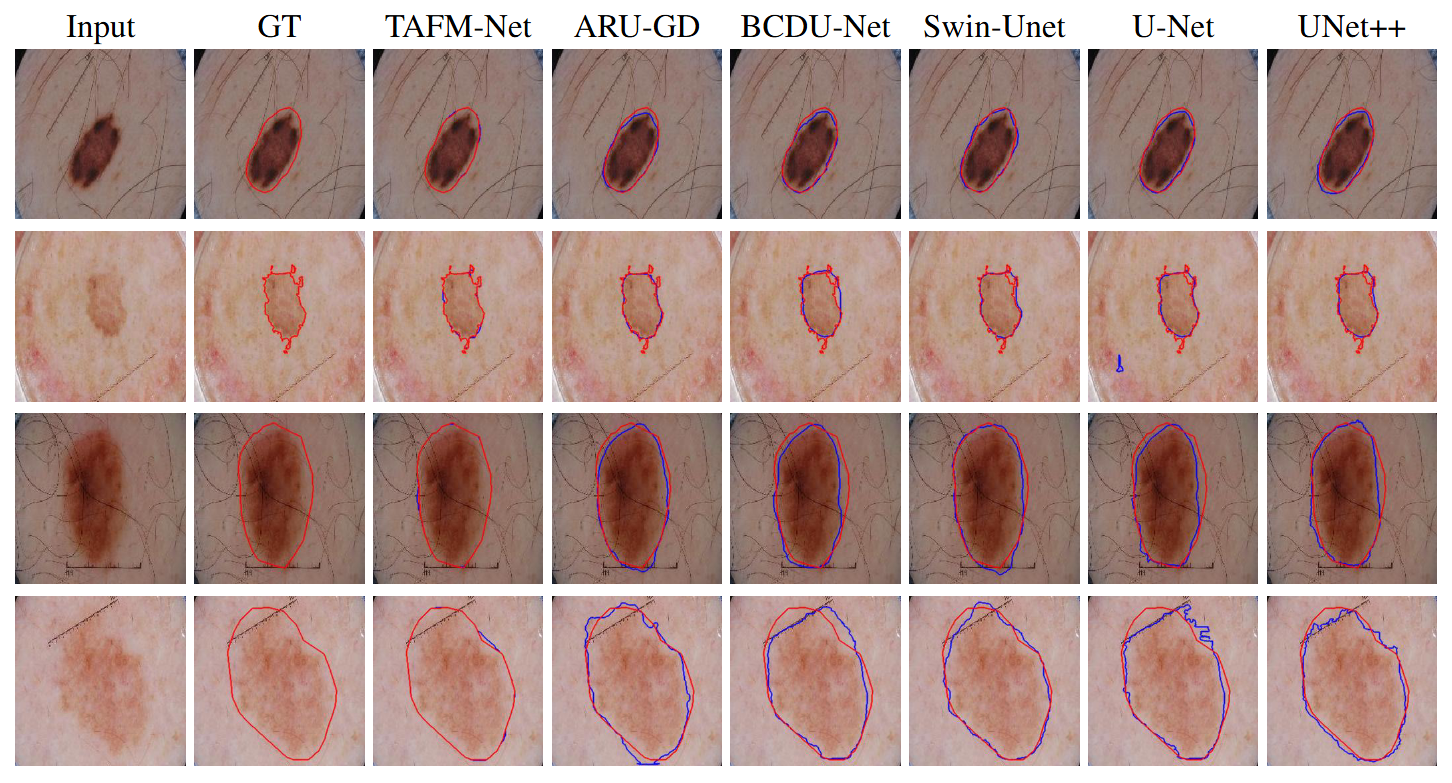}
    \caption{Visual performance comparison of TAFM-Net with other SOTA methods on four example cases from the ISIC 2017 dataset.}
    \label{fig:Vis_ISIC2017}
\end{figure*}

For ISIC 2017 dataset, the results (Table \ref{tab:skin}) show that TAFM-Net outperformed existing methods in almost all cases, achieving improvements in the Jaccard score of 3.8\%--17.2\%. Example visual comparisons (Fig.~\ref{fig:Vis_ISIC2017}) with selected available methods confirm the superiority of TAFM-Net, especially in images with complications such as hairs, low contrast, variations in lesion size, and irregular boundaries.

\begin{figure*}[!t]
    \centering
    \includegraphics[width=\textwidth]{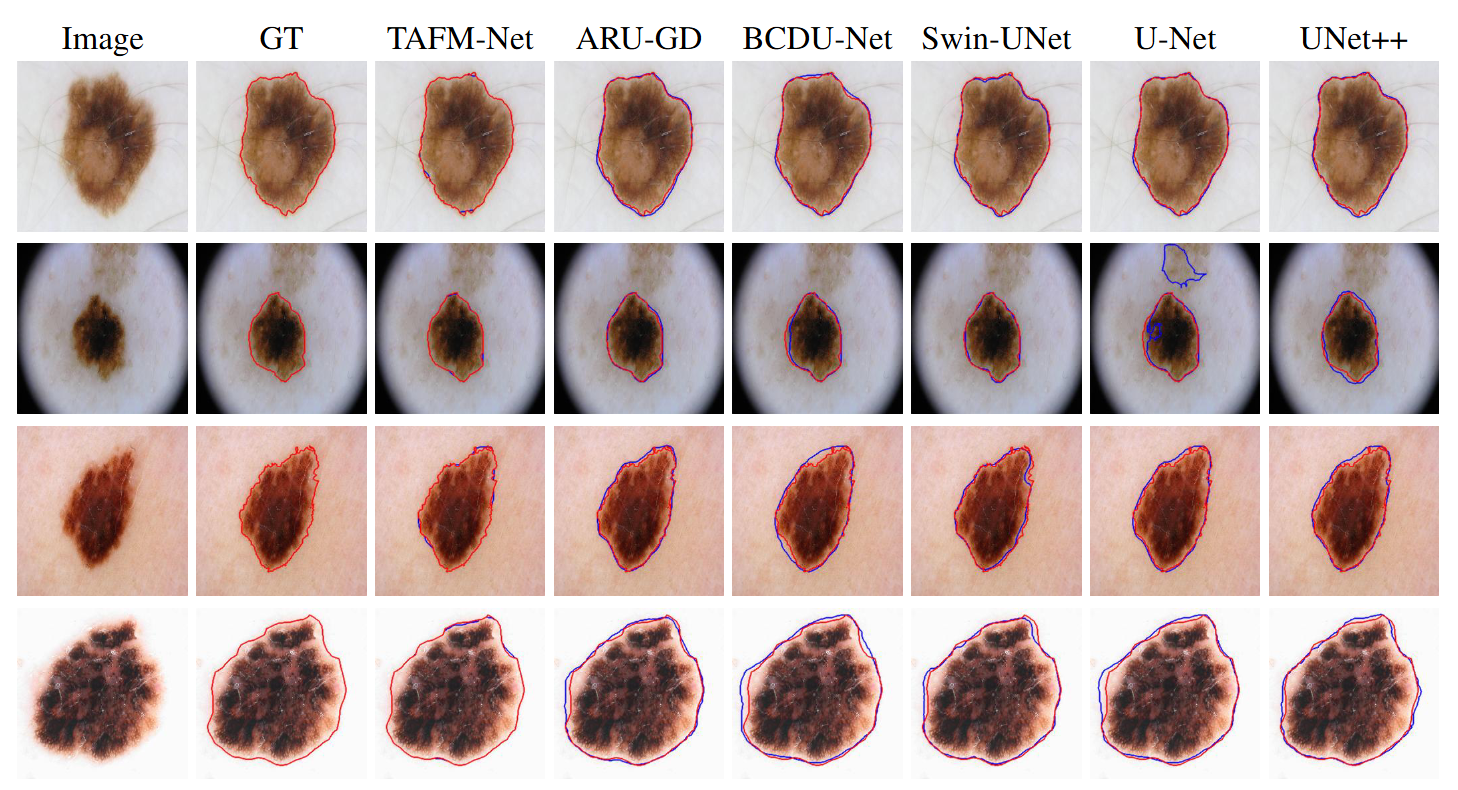}
    \caption{Visual performance comparison of TAFM-Net with other SOTA methods on four example cases from the ISIC 2016 dataset.}
    \label{fig:Vis_ISIC2016}
\end{figure*}

Finally, for ISIC2016 dataset, the results (Table \ref{tab:skin}) show that TAFM-Net consistently outperformed existing methods, achieving improvements in the Jaccard score of 6.2\%--15.1\%. The example visual comparisons (Fig.~\ref{fig:Vis_ISIC2016}) with the available selected methods again confirm the superiority of TAFM-Net, especially in images where the contrast fluctuates, the lesion sizes differ and the boundaries are irregular.

\subsubsection{Comparison on PH2 Dataset}
\begin{figure*}[!t]
    \centering
    \includegraphics[width=\textwidth]{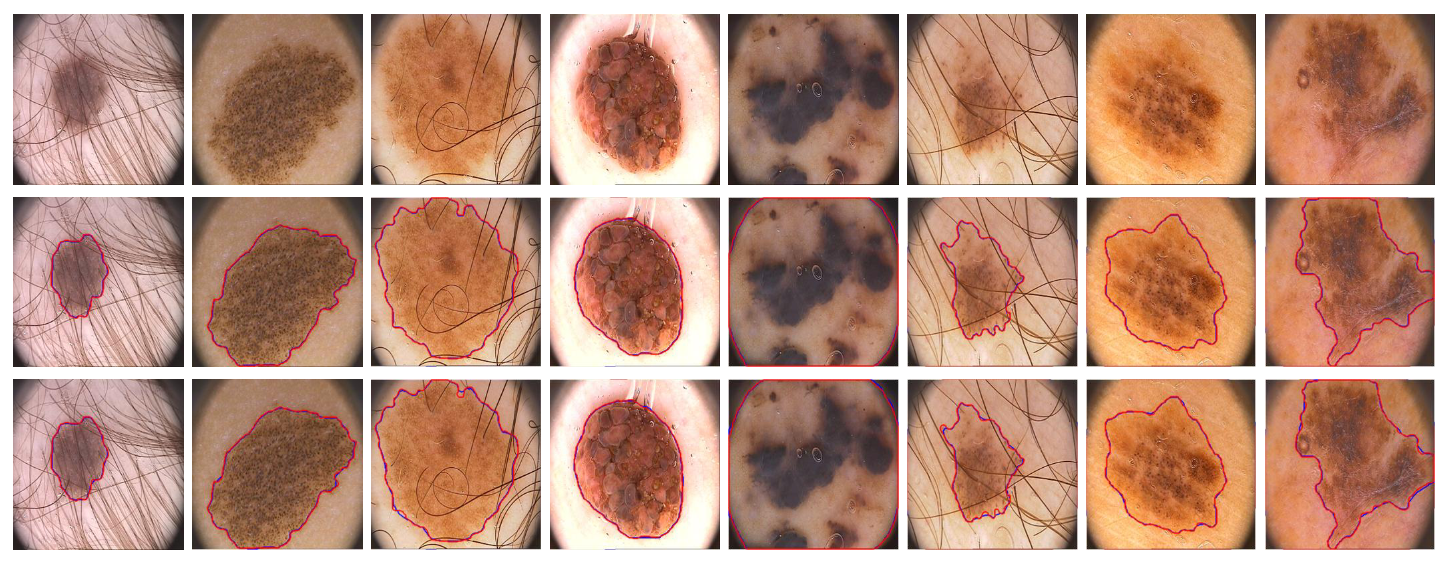}
    \caption{Visual performance of TAFM-Net on eight examples cases from the PH2 dataset with training done on the ISIC 2016 dataset. Rows from top to bottom: input images, corresponding ground-truth images, and outputs of TAFM-Net.}
    \label{fig:Vis_PH2}
\end{figure*}

Generalisation of TAFM-Net was evaluated by cross-data set validation. We trained the network on the ISIC 2016 dataset and tested it on the PH2 dataset. The same was done with the following SOTA comparison methods: AS-Net \cite{HU2022117112}, DCL-PSI \cite{bi2019step}, ICL-Net \cite{cao2022icl}, and MFCN \cite{bi2019step}. The results (Table \ref{tab:PH2}) show that TAFM-Net again consistently outperformed existing methods, achieving improvements in the Jaccard score of 9.1\%--13.8\%. Visual examples of TAFM-Net segmentation results (Fig.~\ref{fig:Vis_PH2}) demonstrate the ability of the model to accurately delineate lesion regions in the context of challenges such as hairs, variations in lesion size, contrast variations, and irregular boundaries.

\begin{table}[!t]
  \centering
  \caption{Performance comparison of the proposed TAFM-Net and existing SOTA methods with training on the ISIC 2016 dataset and testing on the PH2 dataset. Best score per metric in \textbf{bold}.}
  \begin{tabular}{lccccc}
    \toprule
   \multirow{2}[4]{*}{\textbf{Method}} & \multicolumn{5}{c}{\textbf{Performance (\%)}} \\
\cmidrule{2-6} & $\mathbf{A}$ & $\mathbf{Sn}$ & $\mathbf{Sp}$ & $\mathbf{J}$ & $\mathbf{D}$ \\
   \midrule
    AS-Net \cite{HU2022117112} & 95.20 & 96.24 & 94.31 & 87.60 & 93.05 \\
    DCL-PSI \cite{bi2019step} & 95.30 & 96.23 & 94.52 & 85.90 & 92.10 \\
    ICL-Net \cite{cao2022icl} & 96.32 & 95.46 & 97.36 & 87.25 & 92.80 \\
    MFCN \cite{bi2019step} & 94.24 & 94.89 & 93.98 & 83.99 & 90.66 \\
\midrule
    \textbf{Proposed TAFM-Net} & \textbf{98.13} & \textbf{97.48} & \textbf{98.15} & \textbf{95.60} & \textbf{97.81} \\
    \bottomrule
    \end{tabular}
  \label{tab:PH2}
\end{table}

\subsubsection{Comparison of Performance Trend Over Epochs}
\begin{figure}[!t]
    \centering
    \begin{subfigure}
        \centering
        \includegraphics[width=1.05\textwidth]{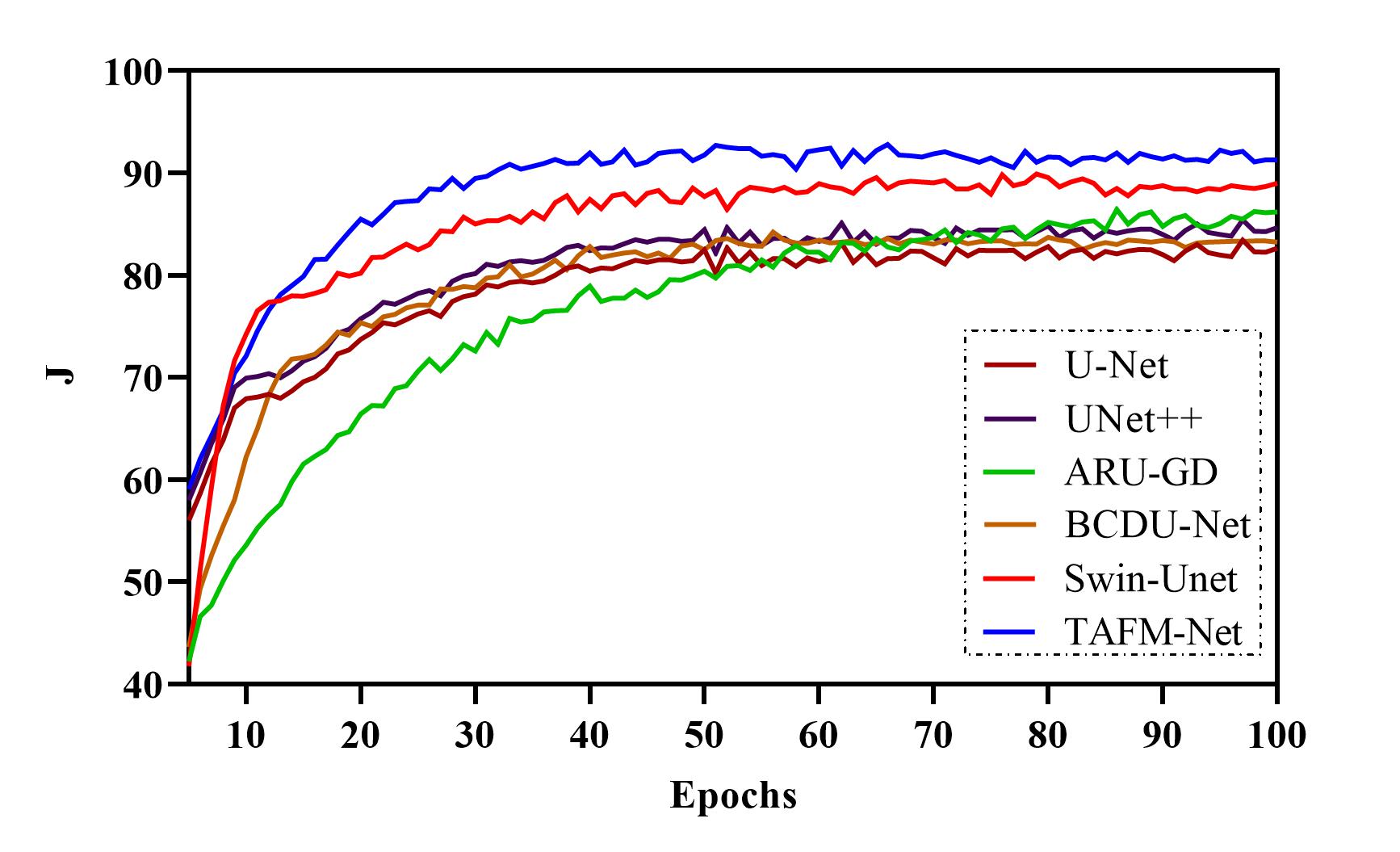}
        \label{fig:jaccard_vs_epoch}
    \end{subfigure}
    \hfill
    \begin{subfigure}
        \centering
        \includegraphics[width=1\textwidth]{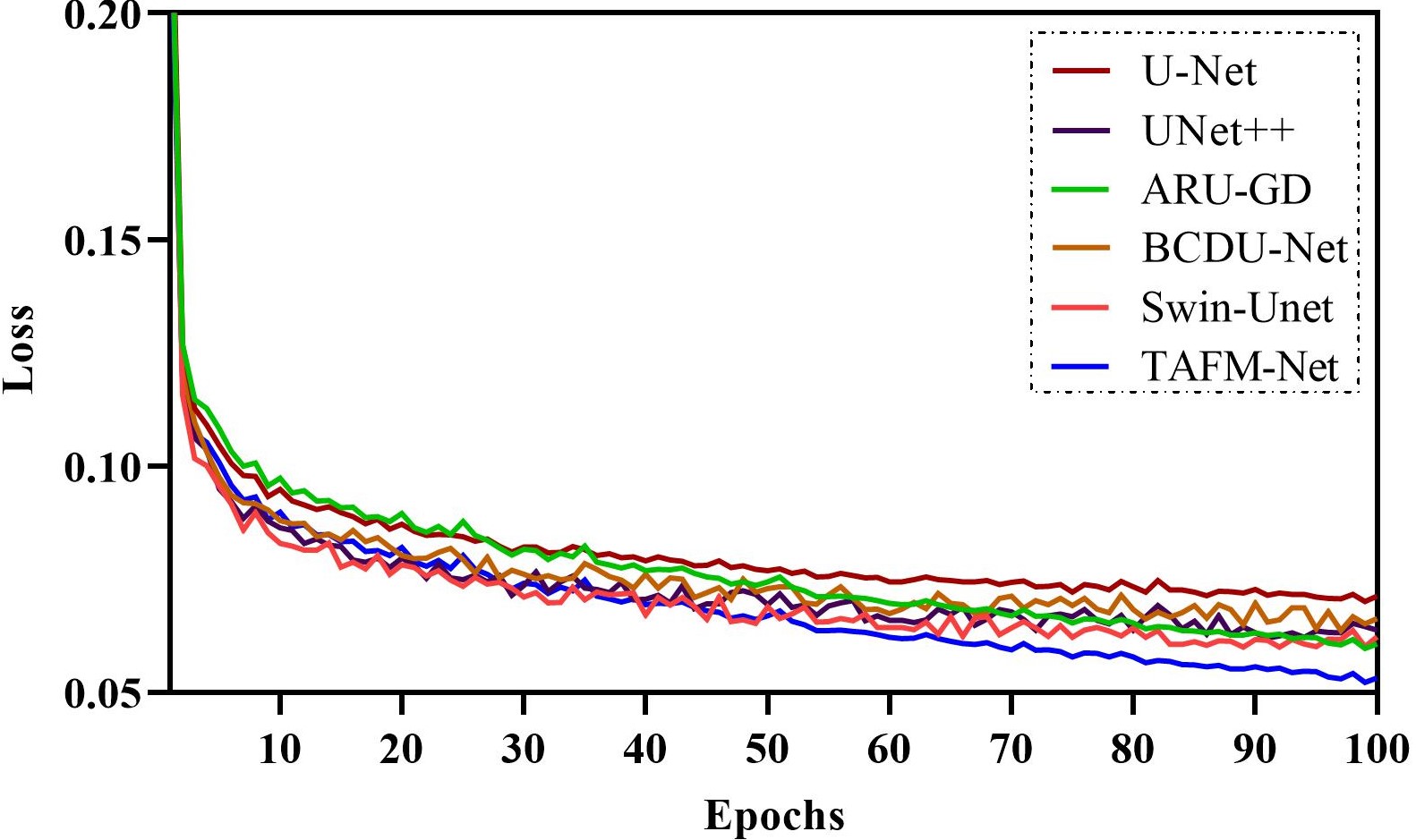}
        \label{fig:loss_vs_epoch}
    \end{subfigure}
    \caption{Performance trend of TAFM-Net and comparison methods on the ISIC 2016 dataset in terms of the Jaccard score (J) and loss as a function of the number of training epochs.}
    \label{fig:performance_trend}
\end{figure}

We also evaluated how TAFM-Net and several SOTA comparison networks improve their performance during training epochs. To do this, throughout the training process using the ISIC 2016 dataset, we tracked the evolution of the Jaccard score (J) and the loss of training data. The results (Fig.~\ref{fig:performance_trend}) show that for TAFM-Net, J crossed 80\% in the 15th epoch, while SwinU-Net crossed that bar in the 20th epoch, and the other comparison networks in much later epochs. The maximum J score of TAFM-Net was 95.32\%, while SwinU-Net achieved a maximum of 89.91\%. Furthermore, the loss converged faster with TAFM-Net than with the other methods.

\subsubsection{Comparison of Computational Aspects}
As a last evaluation, we compared the computational aspects of TAFM-Net with SOTA methods. Specifically, we looked at the inference time per image (speed), the number of parameters (related to memory usage), and the number of floating-point operations per second (efficiency). From the numbers (Table \ref{tab:Computations}) we conclude that TAFM-Net is the network of choice. With its mere 20.6 million (M) parameters, TAFM-Net is considerably faster, smaller, and more computationally efficient than the comparison networks, while at the same time it outperforms them in terms of segmentation fidelity. The compact nature of TAFM-Net facilitates its implementation and use in clinical settings.

\begin{table}
  \centering
  \caption{Analysis of Computational Complexity for the proposed TAFM-Net. This assessment is performed based on a spatial dimension of $256\times 256$.}
    \begin{tabular}{lccc}
    \toprule
    \multirow{2}[4]{*}{\textbf{Method}} & \multicolumn{3}{c}{\textbf{Computational Aspect}} \\
\cmidrule{2-4}          & \textbf{Inference (ms) $\downarrow$} & \textbf{Params (M) $\downarrow$} & \textbf{FLOPs (G) $\downarrow$} \\
    \midrule
    ARU-GD \cite{Maji2022} & 29.49 & 33.30  & 33.93 \\
    BCDU-Net \cite{azad2019bi} & 28.07 & 28.80  & 38.22 \\
    Swin U-Net \cite{cao2023swin} & 25.60 & 29.00 & 25.40  \\
    U-Net \cite{ronneberger2015} & 28.87 & 32.90  & 33.39 \\
    UNet++ \cite{Zhou2018} & 31.30 & 34.90  & 35.60  \\
    \midrule
    \textbf{Proposed TAFM-Net} & \textbf{18.25} & \textbf{20.60} & \textbf{20.50} \\
    \bottomrule
    \end{tabular}%
  \label{tab:Computations}%
\end{table}%

\section{Conclusions}
We have introduced an innovative and powerful architecture designed to address the critical task of skin lesion segmentation. The proposed TAFM-Net leverages the strengths of an EfficientNet-based U-Net and integrates a transformer attention mechanism and focal modulation. The network offers a multifaceted solution to improve the accuracy and reliability of skin lesion segmentation. Its self-adaptive transformer focuses on both spatial and channel-related saliency in the images. Additionally, the network incorporates a densely connected decoder that integrates focal modulations within skip connections, emphasising essential features. Furthermore, TAFM-Net uses a fused loss specifically designed to promote better alignment between input images and ground truth. To evaluate the effectiveness of TAFM-Net, we conducted extensive experiments on publicly available skin lesion datasets, with a particular focus on datasets from the ISIC archive. Our findings show the competitive performance of our proposed network.

Looking ahead, there are exciting avenues for further improvement. A more powerful encoder architecture (multiple encoders, multiple decoders) could potentially enhance our model's performance. However, this would involve exploring larger parameter spaces. Furthermore, the adaptability of our model could be further refined by optimizing the dynamics of weight loss $\alpha$. This dynamic approach could enable more effective use of the boundary loss, particularly when faced with limited training epochs. Notwithstanding potential further improvements, the current version of TAFM-Net already shows superior performance compared to other SOTA methods, and may serve as a new baseline for future research in skin lesion segmentation.


\footnotesize
\begin{thebibliography}{10}
\expandafter\ifx\csname url\endcsname\relax
  \def\url#1{\texttt{#1}}\fi
\expandafter\ifx\csname urlprefix\endcsname\relax\def\urlprefix{URL }\fi
\expandafter\ifx\csname href\endcsname\relax
  \def\href#1#2{#2} \def\path#1{#1}\fi

\bibitem{khan2024esdmr}
T.~M. Khan, S.~S. Naqvi, E.~Meijering, Esdmr-net: A lightweight network with
  expand-squeeze and dual multiscale residual connections for medical image
  segmentation, Engineering Applications of Artificial Intelligence 133 (2024)
  107995.

\bibitem{khan2022leveraging}
T.~M. Khan, S.~S. Naqvi, E.~Meijering, Leveraging image complexity in
  macro-level neural network design for medical image segmentation, Scientific
  Reports 12~(1) (2022) 22286.

\bibitem{khan2022t}
T.~M. Khan, A.~Robles-Kelly, S.~S. Naqvi, T-net: A resource-constrained tiny
  convolutional neural network for medical image segmentation, in: Proceedings
  of the IEEE/CVF winter conference on applications of computer vision, 2022,
  pp. 644--653.

\bibitem{khan2022mkis}
T.~M. Khan, M.~Arsalan, A.~Robles-Kelly, E.~Meijering, Mkis-net: a light-weight
  multi-kernel network for medical image segmentation, in: International
  Conference on Digital Image Computing: Techniques and Applications (DICTA),
  10.1109/DICTA56598.2022.10034573, 2022, pp. 1--8.

\bibitem{naqvi2023glan}
S.~S. Naqvi, Z.~A. Langah, H.~A. Khan, M.~I. Khan, T.~Bashir, M.~I. Razzak,
  T.~M. Khan, Glan: Gan assisted lightweight attention network for biomedical
  imaging based diagnostics, Cognitive Computation 15~(3) (2023) 932--942.

\bibitem{iqbal2023ldmres}
S.~Iqbal, T.~M. Khan, S.~S. Naqvi, A.~Naveed, M.~Usman, H.~A. Khan, I.~Razzak,
  Ldmres-net: A lightweight neural network for efficient medical image
  segmentation on iot and edge devices, IEEE journal of biomedical and health
  informatics (2023).

\bibitem{qayyum2023two}
A.~Qayyum, I.~Razzak, M.~Mazher, T.~Khan, W.~Ding, S.~Niederer, Two-stage
  self-supervised contrastive learning aided transformer for real-time medical
  image segmentation, IEEE Journal of Biomedical and Health Informatics (2023).

\bibitem{javed2024advancing}
S.~Javed, T.~M. Khan, A.~Qayyum, A.~Sowmya, I.~Razzak, Advancing medical image
  segmentation with mini-net: A lightweight solution tailored for efficient
  segmentation of medical images, arXiv preprint arXiv:2405.17520 (2024).

\bibitem{iqbal2022recent}
S.~Iqbal, T.~M. Khan, K.~Naveed, S.~S. Naqvi, S.~J. Nawaz, {Recent trends and
  advances in fundus image analysis: A review}, Compt. in Biology and Medicine
  (2022) 106277.

\bibitem{soomro2016automatic}
T.~A. Soomro, M.~A. Khan, J.~Gao, T.~M. Khan, M.~Paul, N.~Mir, Automatic
  retinal vessel extraction algorithm, in: 2016 International Conference on
  Digital Image Computing: Techniques and Applications (DICTA), IEEE, 2016, pp.
  1--8.

\bibitem{khan2019boosting}
M.~A. Khan, T.~M. Khan, T.~A. Soomro, N.~Mir, J.~Gao, Boosting sensitivity of a
  retinal vessel segmentation algorithm, Pattern Analysis and Applications 22
  (2019) 583--599.

\bibitem{khan2020shallow}
T.~M. Khan, F.~Abdullah, S.~S. Naqvi, M.~Arsalan, M.~A. Khan, Shallow vessel
  segmentation network for automatic retinal vessel segmentation, in: 2020
  International Joint Conference on Neural Networks (IJCNN), IEEE, 2020, pp.
  1--7.

\bibitem{arsalan2022prompt}
M.~Arsalan, T.~M. Khan, S.~S. Naqvi, M.~Nawaz, I.~Razzak, Prompt deep
  light-weight vessel segmentation network (plvs-net), IEEE/ACM Transactions on
  Computational Biology and Bioinformatics 20~(2) (2022) 1363--1371.

\bibitem{khan2022neural}
T.~M. Khan, S.~S. Naqvi, A.~Robles-Kelly, E.~Meijering, Neural network
  compression by joint sparsity promotion and redundancy reduction, in:
  International Conference on Neural Information Processing, Springer
  International Publishing Cham, 2022, pp. 612--623.

\bibitem{khan2023retinal}
T.~M. Khan, S.~S. Naqvi, A.~Robles-Kelly, I.~Razzak, Retinal vessel
  segmentation via a multi-resolution contextual network and adversarial
  learning, Neural Networks 165 (2023) 310--320.

\bibitem{khan2020exploiting}
T.~M. Khan, S.~S. Naqvi, M.~Arsalan, M.~A. Khan, H.~A. Khan, A.~Haider,
  Exploiting residual edge information in deep fully convolutional neural
  networks for retinal vessel segmentation, in: 2020 International Joint
  Conference on Neural Networks (IJCNN), IEEE, 2020, pp. 1--8.

\bibitem{khan2020semantically}
T.~M. Khan, A.~Robles-Kelly, S.~S. Naqvi, A semantically flexible feature
  fusion network for retinal vessel segmentation, in: International Conference
  on Neural Information Processing, Springer, Cham, 2020, pp. 159--167.

\bibitem{khan2021residual}
T.~M. Khan, A.~Robles-Kelly, S.~S. Naqvi, A.~Muhammad, Residual multiscale full
  convolutional network (rm-fcn) for high resolution semantic segmentation of
  retinal vasculature, in: Structural, Syntactic, and Statistical Pattern
  Recognition: Joint IAPR International Workshops, S+ SSPR 2020, Padua, Italy,
  January 21--22, 2021, Proceedings, Springer Nature, 2021, p. 324.

\bibitem{khan2021rc}
T.~M. Khan, A.~Robles-Kelly, S.~S. Naqvi, Rc-net: A convolutional neural
  network for retinal vessel segmentation, in: 2021 Digital Image Computing:
  Techniques and Applications (DICTA), IEEE, 2021, pp. 01--07.

\bibitem{naveed2024pca}
A.~Naveed, S.~S. Naqvi, T.~M. Khan, I.~Razzak, Pca: progressive class-wise
  attention for skin lesions diagnosis, Engineering Applications of Artificial
  Intelligence 127 (2024) 107417.

\bibitem{naveed2024ra}
A.~Naveed, S.~S. Naqvi, S.~Iqbal, I.~Razzak, H.~A. Khan, T.~M. Khan, Ra-net:
  Region-aware attention network for skin lesion segmentation, Cognitive
  Computation (2024) 1--18.

\bibitem{iqbal2024tesl}
S.~Iqbal, M.~Zeeshan, M.~Mehmood, T.~M. Khan, I.~Razzak, Tesl-net: A
  transformer-enhanced cnn for accurate skin lesion segmentation, arXiv
  preprint arXiv:2408.09687 (2024).

\bibitem{naveed2024ad}
A.~Naveed, S.~S. Naqvi, T.~M. Khan, S.~Iqbal, M.~Y. Wani, H.~A. Khan, Ad-net:
  Attention-based dilated convolutional residual network with guided decoder
  for robust skin lesion segmentation, Neural Computing and Applications (2024)
  1--23.

\bibitem{ronneberger2015}
O.~Ronneberger, P.~Fischer, T.~Brox, {U-Net}: Convolutional networks for
  biomedical image segmentation, in: Medical Image Computing and
  Computer-Assisted Intervention (MICCAI), 2015, pp. 234--241.

\bibitem{ghafoorian2016}
M.~Ghafoorian, N.~Karssemeijer, T.~Heskes, I.~W.~M. van Uder, F.~E. de~Leeuw,
  E.~Marchiori, B.~van Ginneken, B.~Platel, Non-uniform patch sampling with
  deep convolutional neural networks for white matter hyperintensity
  segmentation, in: IEEE International Symposium on Biomedical Imaging (ISBI),
  2016, pp. 1414--1417.

\bibitem{yu2017}
L.~Yu, H.~Chen, Q.~Dou, J.~Qin, P.-A. Heng, Automated melanoma recognition in
  dermoscopy images via very deep residual networks, IEEE Transactions on
  Medical Imaging 36~(4) (2017) 994--1004.

\bibitem{BASAK2022108673}
H.~Basak, R.~Kundu, R.~Sarkar, {MFSNet: A multi focus segmentation network for
  skin lesion segmentation}, Pattern Recognition 128 (2022) 108673.

\bibitem{WANG2022108636}
K.~Wang, X.~Zhang, X.~Zhang, Y.~Lu, S.~Huang, D.~Yang, Eanet: Iterative edge
  attention network for medical image segmentation, Pattern Recognition 127
  (2022) 108636.

\bibitem{oktay2018}
O.~Oktay, J.~Schlemper, L.~L. Folgoc, M.~Lee, M.~Heinrich, K.~Misawa, K.~Mori,
  S.~McDonagh, N.~Y. Hammerla, B.~Kainz, B.~Glocker, D.~Rueckert, {Attention
  U-Net}: Learning where to look for the pancreas, arXiv:1804.03999 (2018).

\bibitem{zhang2019}
J.~Zhang, Y.~Xie, Y.~Xia, C.~Shen, Attention residual learning for skin lesion
  classification, IEEE Transactions on Medical Imaging 38~(9) (2019)
  2092--2103.

\bibitem{woo2018}
S.~Woo, J.~Park, J.-Y. Lee, I.~S. Kweon, {CBAM: Convolutional block attention
  module}, in: European Conference on Computer Vision (ECCV), 2018, pp. 3--19.

\bibitem{farooq2024lssf}
H.~Farooq, Z.~Zafar, A.~Saadat, T.~M. Khan, S.~Iqbal, I.~Razzak, Lssf-net:
  Lightweight segmentation with self-awareness, spatial attention, and focal
  modulation, arXiv preprint arXiv:2409.01572 (2024).

\bibitem{iqbal2025tbconvl}
S.~Iqbal, T.~M. Khan, S.~S. Naqvi, A.~Naveed, E.~Meijering, Tbconvl-net: A
  hybrid deep learning architecture for robust medical image segmentation,
  Pattern Recognition 158 (2025) 111028.

\bibitem{khan2024lmbf}
T.~M. Khan, S.~Iqbal, S.~S. Naqvi, I.~Razzak, E.~Meijering, Lmbf-net: A
  lightweight multipath bidirectional focal attention network for multifeatures
  segmentation, in: 2024 IEEE International Conference on Image Processing
  (ICIP), IEEE, 2024, pp. 2807--2813.

\bibitem{jiang2022seacu}
X.~Jiang, J.~Jiang, B.~Wang, J.~Yu, J.~Wang, {SEACU-Net: Attentive ConvLSTM
  U-Net with squeeze-and-excitation layer for skin lesion segmentation},
  Computer Methods and Programs in Biomedicine 225 (2022) 107076.

\bibitem{azad2019bi}
R.~Azad, M.~Asadi-Aghbolaghi, M.~Fathy, S.~Escalera, Bi-directional {ConvLSTM
  U-Net} with densely connected convolutions, in: IEEE/CVF International
  Conference on Computer Vision Workshops (ICCVW), 2019, pp. 406--415.

\bibitem{song2018pyramid}
H.~Song, W.~Wang, S.~Zhao, J.~Shen, K.-M. Lam, Pyramid dilated deeper
  {ConvLSTM} for video salient object detection, in: European Conference on
  Computer Vision (ECCV), 2018, pp. 715--731.

\bibitem{Zhou2018}
Z.~Zhou, M.~M. Rahman~Siddiquee, N.~Tajbakhsh, J.~Liang, {UNet++}: A nested
  {U-Net} architecture for medical image segmentation, in: Deep Learning in
  Medical Image Analysis and Multimodal Learning for Clinical Decision Support,
  2018, pp. 3--11.

\bibitem{Maji2022}
D.~Maji, P.~Sigedar, M.~Singh, {Attention Res-UNet} with guided decoder for
  semantic segmentation of brain tumors, Biomedical Signal Processing and
  Control 71 (2022) 103077.

\bibitem{Aghdam2022}
E.~K. Aghdam, R.~Azad, M.~Zarvani, D.~Merhof, {Attention Swin U-Net}:
  Cross-contextual attention mechanism for skin lesion segmentation,
  arXiv:2210.16898 (2022).

\bibitem{cao2023swin}
H.~Cao, Y.~Wang, J.~Chen, D.~Jiang, X.~Zhang, Q.~Tian, M.~Wang, {Swin-Unet}:
  {Unet}-like pure transformer for medical image segmentation, in: European
  Conference on Computer Vision Workshops (ECCVW), 2023, pp. 205--218.

\bibitem{FIAZ2024110812}
M.~Fiaz, M.~Noman, H.~Cholakkal, R.~M. Anwer, J.~Hanna, F.~S. Khan,
  Guided-attention and gated-aggregation network for medical image
  segmentation, Pattern Recognition 156 (2024) 110812.

\bibitem{Chen2021}
B.~Chen, Y.~Liu, Z.~Zhang, G.~Lu, A.~W.~K. Kong, {TransAttUnet}: Multi-level
  attention-guided {U-Net} with {Transformer} for medical image segmentation,
  IEEE Transactions on Emerging Topics in Computational Intelligence 8~(1)
  (2024) 55--68.

\bibitem{HUANG2024110375}
Z.~Huang, S.~Cheng, L.~Wang, Medical image segmentation based on dynamic
  positioning and region-aware attention, Pattern Recognition 151 (2024)
  110375.

\bibitem{Dong2022}
Y.~Dong, L.~Wang, Y.~Li, {TC-Net: Dual coding network of Transformer and CNN
  for skin lesion segmentation}, PLoS One 17~(11) (2022) e0277578.

\bibitem{Feng2022}
K.~Feng, L.~Ren, G.~Wang, H.~Wang, Y.~Li, {SLT-Net: A codec network for skin
  lesion segmentation}, Computers in Biology and Medicine 148 (2022) 105942.

\bibitem{YUAN2023109228}
F.~Yuan, Z.~Zhang, Z.~Fang, An effective cnn and transformer complementary
  network for medical image segmentation, Pattern Recognition 136 (2023)
  109228.

\bibitem{GUO2024110491}
X.~Guo, X.~Lin, X.~Yang, L.~Yu, K.-T. Cheng, Z.~Yan, Uctnet: Uncertainty-guided
  cnn-transformer hybrid networks for medical image segmentation, Pattern
  Recognition 152 (2024) 110491.

\bibitem{Tan2021}
M.~Tan, Q.~V. Le, {EfficientNetV2}: Smaller models and faster training,
  arXiv:2104.00298 (2021).

\bibitem{naderi2022focal}
M.~Naderi, M.~Givkashi, F.~Piri, N.~Karimi, S.~Samavi, {Focal-UNet: UNet}-like
  focal modulation for medical image segmentation, arXiv:2212.09263 (2022).

\bibitem{Abraham2018}
N.~Abraham, N.~M. Khan, A novel focal {Tversky} loss function with improved
  attention {U-Net} for lesion segmentation, arXiv:1810.07842 (2018).

\bibitem{Kervadec2021}
H.~Kervadec, J.~Bouchtiba, C.~Desrosiers, E.~Granger, J.~Dolz, I.~{Ben Ayed},
  {Boundary loss for highly unbalanced segmentation}, Medical Image Analysis 67
  (2021) 101851.

\bibitem{Mirikharaji2022}
Z.~Mirikharaji, K.~Abhishek, A.~Bissoto, C.~Barata, S.~Avila, E.~Valle, M.~E.
  Celebi, G.~Hamarneh, A survey on deep learning for skin lesion segmentation,
  Medical Image Analysis 88 (2023) 102863.

\bibitem{selvaraju2017grad}
R.~R. Selvaraju, M.~Cogswell, A.~Das, R.~Vedantam, D.~Parikh, D.~Batra,
  {Grad-CAM}: Visual explanations from deep networks via gradient-based
  localization, in: IEEE International Conference on Computer Vision (ICCV),
  2017, pp. 618--626.

\bibitem{yang2022focal}
J.~Yang, C.~Li, X.~Dai, J.~Gao, Focal modulation networks, Advances in Neural
  Information Processing Systems 35 (2022) 4203--4217.

\bibitem{wang2023autosmim}
Z.~Wang, J.~Lyu, X.~Tang, {autoSMIM}: Automatic superpixel-based masked image
  modeling for skin lesion segmentation, IEEE Transactions on Medical Imaging
  42~(12) (2023) 3501--3511.

\bibitem{9049412}
S.~Feng, H.~Zhao, F.~Shi, X.~Cheng, M.~Wang, Y.~Ma, D.~Xiang, W.~Zhu, X.~Chen,
  {CPFNet}: Context pyramid fusion network for medical image segmentation, IEEE
  Transactions on Medical Imaging 39~(10) (2020) 3008--3018.

\bibitem{Lei2020}
B.~Lei, Z.~Xia, F.~Jiang, X.~Jiang, Z.~Ge, Y.~Xu, J.~Qin, S.~Chen, T.~Wang,
  S.~Wang, {Skin lesion segmentation via generative adversarial networks with
  dual discriminators}, Medical Image Analysis 64 (2020) 101716.

\bibitem{bi2022hyper}
L.~Bi, M.~Fulham, J.~Kim, {Hyper-fusion network for semi-automatic segmentation
  of skin lesions}, Medical Image Analysis 76 (2022) 102334.

\bibitem{HU2022117112}
K.~Hu, J.~Lu, D.~Lee, D.~Xiong, Z.~Chen, {AS-Net}: Attention synergy network
  for skin lesion segmentation, Expert Systems with Applications 201 (2022)
  117112.

\bibitem{bi2019step}
L.~Bi, J.~Kim, E.~Ahn, A.~Kumar, D.~Feng, M.~Fulham, {Step-wise integration of
  deep class-specific learning for dermoscopic image segmentation}, Pattern
  Recognition 85 (2019) 78--89.

\bibitem{cao2022icl}
W.~Cao, G.~Yuan, Q.~Liu, C.~Peng, J.~Xie, X.~Yang, X.~Ni, J.~Zheng, {ICL-Net}:
  Global and local inter-pixel correlations learning network for skin lesion
  segmentation, IEEE Journal of Biomedical and Health Informatics 27~(1) (2023)
  145--156.

\end{thebibliography}
\end{document}